\documentclass{aa} 
\usepackage{amsmath}
\usepackage{txfonts} 
\usepackage{graphicx}
\usepackage{subfigure}
\usepackage{color}
\usepackage{bm}
\usepackage{url}
\usepackage[]{natbib}
\def\ltsima{$\; \buildrel < \over \sim \;$}
\def\simlt{\lower.5ex\hbox{\ltsima}}
\begin{document} 
\title{ Probing the role of protostellar feedback in clustered star formation\thanks{Based on observations carried out with the IRAM~30m telescope. 
IRAM is supported by INSU/CNRS (France), MPG (Germany), and IGN (Spain).}$^{,}$
\thanks{$^{12}$CO(2--1) data cubes in FITS format are available at the CDS via anonymous ftp to cdsarc.u-strasbg.fr (130.79.128.5) or via http://cdsweb.u-strasbg.fr}}
\subtitle{Mapping outflows in the collapsing protocluster NGC~2264-C} 
\author{A. J. Maury\inst{1} 
\and Ph. Andr\'e\inst{1} 
\and Z.-Y. Li\inst{2}} 
\institute{Laboratoire AIM, CEA/DSM-CNRS-Universit\'e Paris Diderot, IRFU/Service d'Astrophysique, C.E. Saclay, 
Orme des Merisiers, 91191 Gif-sur-Yvette, France
\and Department of Astronomy, University of Virginia, P.O. Box 400325, Charlottesville, VA 22904, USA} 
\date{Received 28 November 2008 / Accepted --} 
\abstract 
{The role played by protostellar feedback in clustered star formation is still a matter of debate.
In particular, protostellar outflows have been proposed as a source of turbulence in cluster-forming clumps, 
which may provide support against global collapse for several free-fall times.} 
{Here, we seek to test the above hypothesis in the case of the well-documented NGC~2264-C protocluster, by quantifying the amount of turbulence and support injected in the surrounding medium by protostellar 
outflows.} 
{Using the HERA heterodyne array on the IRAM 30m telescope, we carried out an extensive mapping of NGC~2264-C 
in the three molecular line transitions $^{12}$CO(2--1), $^{13}$CO(2--1), and C$^{18}$O(2--1).} 
{We found widespread high-velocity  $^{12}$CO emission, testifying to the presence of eleven outflow lobes, closely 
linked to the compact millimeter continuum sources previously detected in the protocluster. 
We carried out a detailed analysis of the dynamical parameters of these outflows, including a quantitative evaluation 
of the overall momentum flux injected in the cluster-forming clump. These dynamical parameters were compared 
to the gravitational and turbulent properties of the clump.}
{We show that the population of protostellar outflows identified in NGC~2264-C are likely to contribute a significant 
fraction of the observed turbulence but cannot efficiently support the protocluster against global collapse.
Gravity appears to largely dominate the dynamics of the NGC~2264-C clump at the present time.
It is however possible that an increase in the star formation rate during the further evolution of the protocluster 
will trigger sufficient outflows to finally halt the contraction of the cloud.} 
\keywords{stars : formation : circumstellar matter  -- ISM : individual objects : NGC~2264-C  -- ISM : kinematics and dynamics  -- ISM : outflows}
\maketitle

\section{Introduction} 
\subsection{Clustered star formation, turbulence and protostellar feedback}

~~~~ It is now well established that a large fraction of young stars in giant molecular clouds form in groups and clusters rather 
than in isolation (e.g. \citealt{carpenter2000,adams2001,lada2003}).

Three main classes of models have been proposed for regulating the star formation process and explaining the origin of
the stellar initial mass function (IMF) in young star clusters. 
The first scenario is based on turbulent fragmentation of the parent molecular cloud (e.g. \citealt{padoan2002,hennebelle2008}), 
which produces an IMF-like core mass distribution as observed in the nearest cluster-forming regions (e.g. \citealt{motte1998}, \citealt{stanke2006}).  
Briefly, self-gravitating pre-stellar condensations 
(each containing one local Jeans mass) form as turbulence-generated density fluctuations, then decouple from their 
turbulent environment, and eventually collapse with little interaction with their surroundings. 
In this picture, the IMF results primarily from the properties (e.g. power spectrum) of {\it interstellar} turbulence, which also 
controls the global star formation rate and efficiency (e.g. \citealt{maclow2004}, \citealt{krumholz2007}).
The second class of models emphasizes the role of {\it protostellar} turbulence and feedback 
in regulating the star formation process (e.g. \citealt{norman1980,adams1996}, \citealt{shu2004}). 
Here, the IMF and the star formation efficiency are determined by the stars themselves through the collective effects of their feedback 
on both individual cores and the parent cloud.
A third scenario exists, however, according to which turbulence is only responsible for forming protostellar seeds but 
plays no direct role in regulating the star formation process and shaping the IMF, 
which is mainly determined by competitive accretion 
and dynamical interactions between individual cluster members (e.g. \citealt{bonnell2001,bate2003}).

From an observational perspective, millimeter-wave observations of molecular clouds have 
revealed supersonic linewidths, which are presumably due to turbulent motions. 
Theory suggests that turbulent motions can be treated as an additional pressure (e.g. \citealt{chandra1951,weizsacker1951}), so that supersonic turbulence increases the effective Jeans mass supported against collapse.

Protostellar outflows have long been considered a plausible way of driving a significant fraction of molecular cloud turbulence, 
especially in cluster-forming clumps, where stars form more efficiently than in the bulk of molecular gas 
(\citealt{bally1996,reipurth2001,knee2000}).
Recently, \citet{nakamura2007} discussed the possible effects of protostellar outflows on cluster formation. 
In particular, they argued that, due to its short decay time (e.g. \citealt{maclow1998}), 
the "interstellar turbulence" initially present in a cluster-forming clump is quickly replaced by turbulent motions generated by protostellar outflows.
According to \citeauthor{nakamura2007}, the protostellar outflow-driven turbulence dominates for most of a 
protocluster's lifetime and acts to maintain the cluster-forming region close to overall virial equilibrium 
for several dynamical times, avoiding global free-fall collapse. 
The exact mechanism by which outflows inject turbulence in the parent cloud is not well understood yet 
and may involved fossil cavities (e.g. \citealt{quillen2005, cunningham2006}).
On the scale of an entire giant molecular cloud, the shocks produced by outflows from young stars may not 
inject momentum and energy at a high enough rate to counter the rate at which turbulent energy decays 
(see \citealt{banerjee2007}). 
On the other hand, intense outflow activity within a young protocluster may be sufficient to support the
parsec-scale parent gas clump, thereby modifying the outcome of the star formation process.

As the role of protostellar feedback in cluster-forming clouds is still a matter of debate, detailed studies of 
the dynamical effects of protostellar outflows in young protoclusters, where outflows are particularly strong and numerous, are 
required to fully understand the process of clustered star formation.
In this paper, we test some of the above ideas by quantifying the impact of  protostellar outflows on the dynamics of 
the NGC~2264-C clump, a protocluster at a very early stage of evolution. 

\subsection{Our target region : NGC2264-C}

The NGC~2264 cluster-forming region is located in the Mon~OB1 giant molecular cloud complex at a distance $d \sim 800$~pc.
The cloud associated with NGC~2264 has been the target of several millimeter line studies, including 
an unbiased CO (J=1$\rightarrow$0) survey for molecular outflows \citep{margulis1988} 
and a search for dense gas via multitransitional CS observations (\citealt{wolfchase1995,wolfchasewalker1995}). 
These early studies revealed two prominent molecular clumps, named NGC~2264-C and NGC~2264-D,  
each associated with a CO outflow. 
The NGC~2264 region also harbors a young star cluster of more than 360 near-IR sources (\citealt{lada1993,lada2003}). 
The most luminous object of the cluster, hereafter called IRS1 after Allen (1972), is embedded in NGC~2264-C and associated with the Class~I IRAS source 
IRAS 06384+0932 ($L_{bol} \sim 2300\ L_\odot $ -- Margulis et al. 1989).

\begin{figure}[!ht]
\begin{center}
  \subfigure{\includegraphics[width=0.8\columnwidth,angle=-90,trim=0cm 1.6cm 0cm 5cm,clip=true]{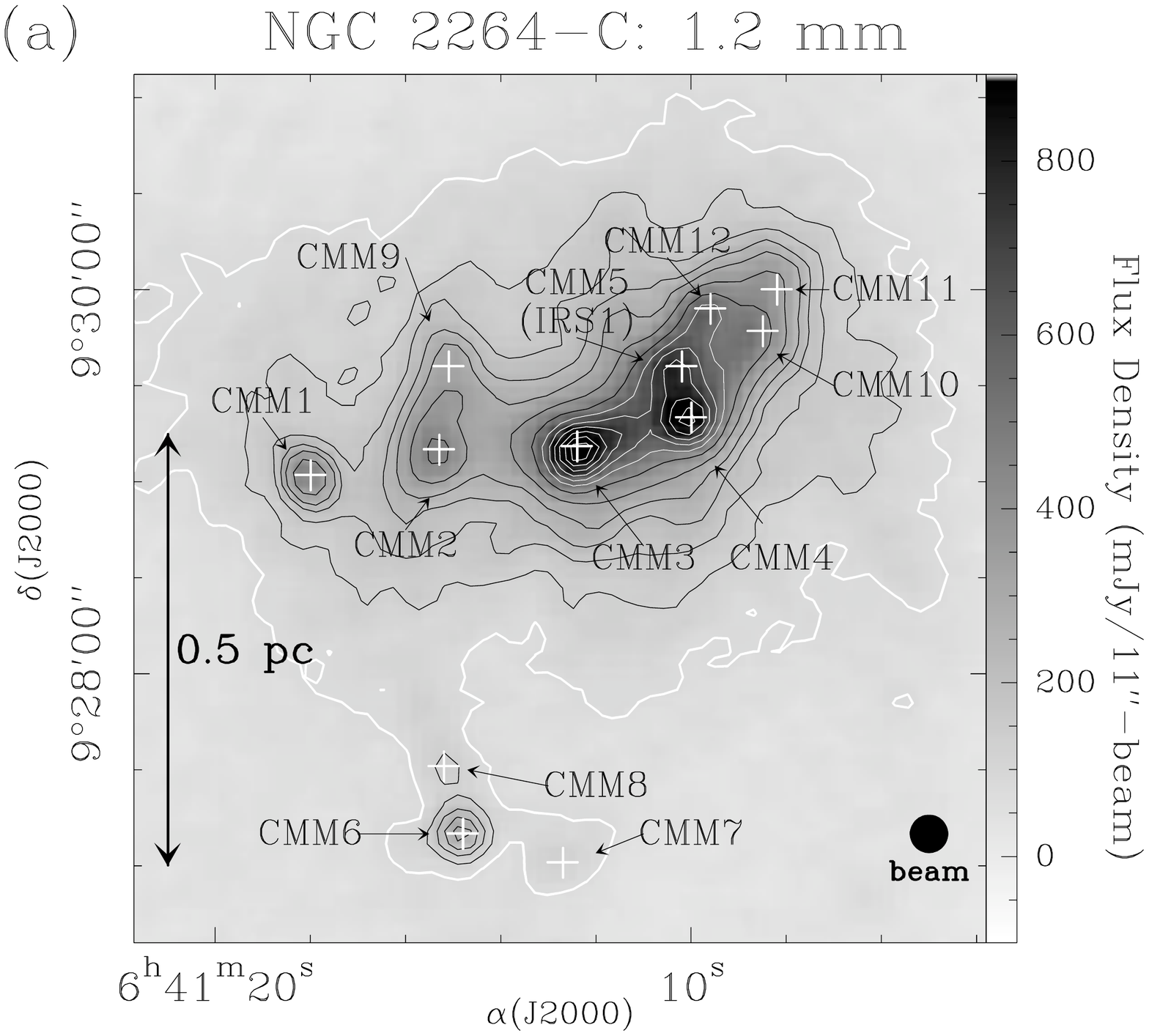}}
  \\
  \subfigure{\includegraphics[width=0.8\columnwidth,angle=-90,trim=0cm 1.6cm 0cm 5cm,clip=true]{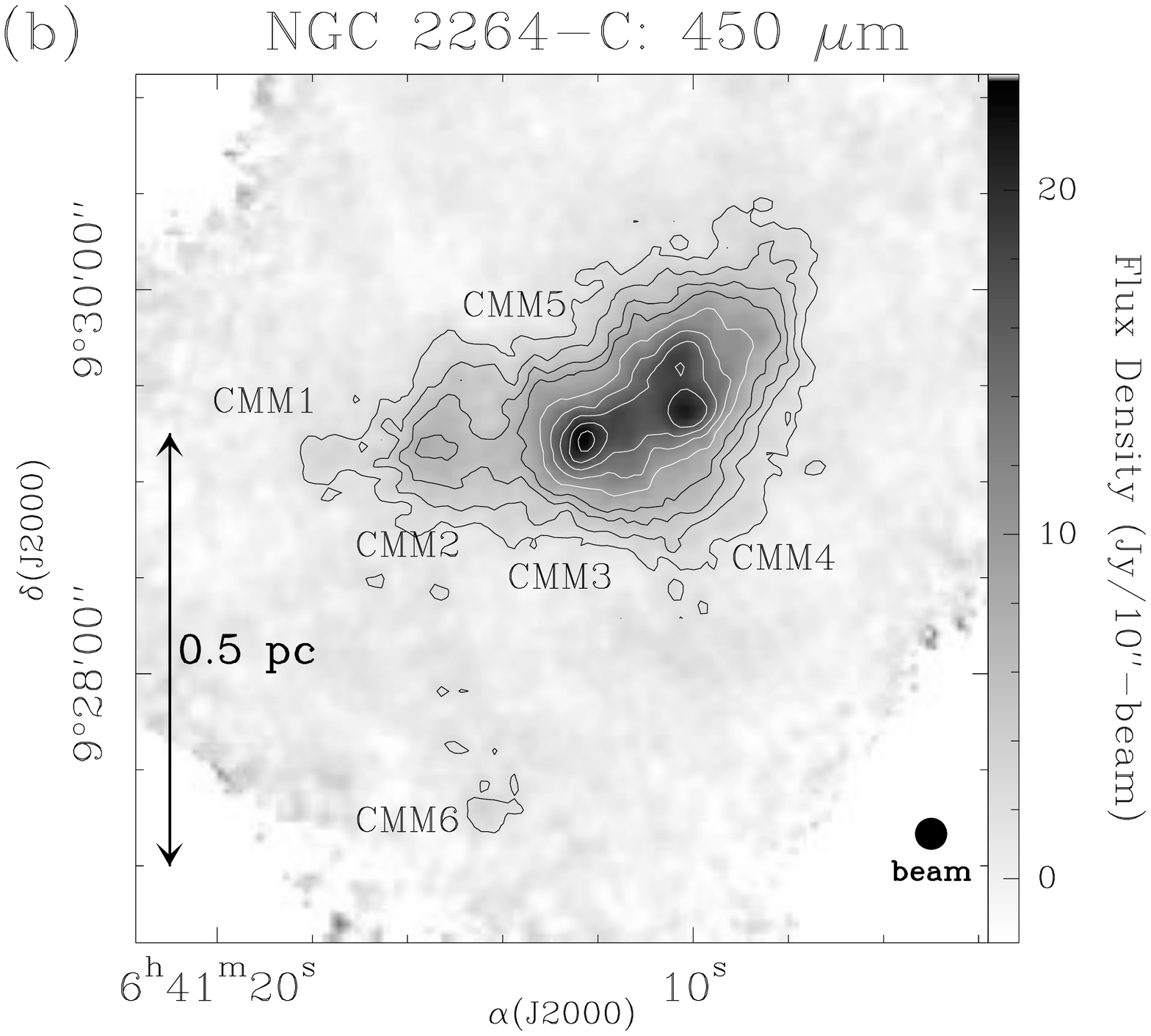}}
\caption{
(a) Dust continuum map of NGC~2264-C taken at 1.2~mm by \citet{peretto2006} 
with the MAMBO bolometer array on the IRAM~30m telescope. 
Twelve millimeter peaks were found and interpreted as candidate Class~0 objects (see Peretto et al. for details). 
(b) Total-power 450~$\mu$m dust continuum map of the NGC~2264-C protocluster taken by us with the 
P-ArT\'eMiS bolometer camera\protect\footnotemark[1]  on the APEX telescope. The HPBW resolution is $\sim 10\arcsec $.
The contour levels go from 1.75 to 7 Jy/10\arcsec -beam by 1.75 Jy/10\arcsec -beam and 
from 10 to 22 Jy/10\arcsec -beam by 3 Jy/10\arcsec -beam.}
\label{fig:cont}
\end{center}
\end{figure}

The internal structure of NGC~2264-C and NGC~2264-D was subsequently resolved by  higher-resolution 
millimeter and submillimeter studies 
(\citealt{wardthompson2000,williams2002,peretto2006,peretto2007}). 
In particular, through 1.2~mm continuum observations, 
\citet{peretto2006} identified 12 compact dense cores 
in NGC~2264-C (see Fig.\ref{fig:cont}), of which at least 8 are Class~0-like objects with associated near-IR H$_2$ jets 
(\citealt{wang2002}) or compact CS outflows (\citealt{schreyer2003}). These cores have typical diameters 
$\sim 0.04$~pc, masses ranging from $\sim 2$ to $\sim 40\, M_{\odot}$, and column densities ranging 
from $\sim 4 \times 10^{22}$ to $\sim 6 \times 10^{23}\, \rm {cm}^{-2}$. Based on HCO$^+$, CS, and N$_2$H$^+$ 
mapping observations, \citet{peretto2006} also showed the 
presence of large-scale collapse motions converging onto the most massive core (C-MM3 with $M \sim 40\, M_{\odot}$), 
near the center of NGC~2264-C.
The IRAM Plateau de Bure interferometer observations of \citet{peretto2007} {\bf revealed} 
a thirteenth source, C-MM13, located only {\bf 13.5$\arcsec$} (or $\sim 0.05$~pc) away from C-MM3. 

\footnotetext[1]{P-ArT\'eMiS is a $16 \times 16$-pixel prototype of the large-format submillimeter bolometer camera 
currently being built by CEA for the APEX telescope (cf. \citet{andre2008,talvard2008}).\\
\textcolor{blue}{\url{http://www.apex-telescope.org/instruments/pi/artemis/}}}

In their study, \citet{peretto2007} sucessfully modeled the dynamical evolution of the 
NGC~2264-C protocluster. 
They compared the available millimeter observations with SPH numerical simulations 
of the collapse of a Jeans-unstable, prolate dense clump. 
\citet{peretto2007} obtained a good quantitative agreement with observations when starting from a low level of initial turbulent energy.
However,  the total mass of dense gas (i.e., with  $ n_{\rm{H_2}} > 10^{4}\, \rm{cm}^{-3}$) was $\sim10$ times lower 
in the best-fit simulations than in the actual NGC~2264-C clump. 
When \citet{peretto2007} increased the mass of dense gas by a factor of 10 in their numerical simulations, a larger 
number of cores was generated, leading to a poor match to the observations. 
It seemed therefore that an additional source of support against gravity, not included in the simulations of \citet{peretto2007}, 
{\bf had to be invoked to explain the current dynamical state NGC 2264-C}. 
Peretto et al. suggested that this extra support could arise from protostellar feedback and/or magnetic fields.

~
\newline\noindent Because it is well documented and known to exhibit outflow activity (\citealt{margulis1988}), the NGC~2264-C protocluster is 
an ideal laboratory for 
probing the initial conditions of clustered star formation and evaluating the impact of outflow feedback on early protocluster evolution. 
We thus initiated a mapping study of the outflow already detected by \citet{margulis1988} in NGC~2264-C, with higher angular resolution and better sensitivity.
Our goal was to assess the momentum injection rate due to outflows in this protocluster and examine whether outflows could affect the global dynamical evolution of the protocluster.
\\ ~~

\section{Observations and data reduction} 

\subsection{Observations} 

Observations of the $^{12}$CO(2--1), C$^{18}$O(2--1) and $^{13}$CO(2--1) emission lines (see Table \ref{tab:vespa_tab}) from the NGC~2264-C protocluster were taken with the IRAM-30 m telescope between October and November 2006 using the HEterodyne Receiver Array HERA (\citealt{schuster2004}) together with the VESPA autocorrelator backend. 
HERA is an assembly of two focal plane arrays (HERA1 and HERA2) of 9 SIS receivers each. The two arrays have orthogonal polarization and their 9 pixels are arranged in the form of a center-filled square and are separated by $24^{\prime\prime}$. 
HERA1 and HERA2 were used simultaneously, allowing us to map the same region of the sky at two different frequencies at the same time.
VESPA was used with a channel spacing of 320 kHz for $^{12}$CO(2--1), and 80kHz for C$^{18}$O(2--1) and $^{13}$CO(2--1). 
The VESPA autocorrelator configurations used during our observations can be found in Table \ref{tab:vespa_tab}.
Observations were carried out in position switched On-The-Fly (OTF) mode, scanning NGC~2264-C in right ascension and declination. We adopted a drift speed of $2\arcsec$/sec with a dump time of 2 s, and HERA was rotated by $9.5\,^{\circ}$ to obtain a spacing of $4\arcsec$ between adjacent scan lines, while a derotator 
was used to keep the HERA pixel pattern stationary in the Nasmyth focal plane. All this resulted in a sampling of $4\arcsec$ over the entire map. 
The half power beamwidth (HPBW) of the 30-m telescope being of 11\arcsec ~at this frequency, this corresponds to an over-Nyquist sampling.
The resulting map has a size of $3.3\arcmin \times 3.3\arcmin$ (equivalent to $\sim0.8$ pc$^2$ at the distance of the protocluster). 
The emission-free reference position 
was located at offsets ($-$1500\arcsec, $-$1500\arcsec) from the protostellar core C-MM3, located at the center of our map. 
The typical system temperatures varied between $\sim $300 K and $\sim $600 K. 
The telescope pointing was checked every $\sim$2 hours on J0528+134, while focus checks and adjustments were made every $\sim$3 hours on the strong source NGC7538. 
The observations were carried out in single sideband mode with an image rejection of $\sim$ 10 dB, providing a calibration accuracy better than $\sim 10\%$.
At 230 GHz, the main beam efficiency and the forward efficiency of the telescope are B$_{\rm{eff}}=0.90$ and F$_{\rm{eff}}=0.52$ respectively. Throughout this paper, the line intensity scale adopted is in units of $T^{*}_{\rm A}$, the equivalent antenna temperature above the atmosphere. 

\begin{table}[!h]
\centering \par \caption{Parameters of the IRAM 30m observations} 
\begin{tabular}{lcccccc|} 
\hline
\hline
{Line} & {Frequency} & {Resolution} & {Bandwidth} & {Mean r.m.s.} \\
{} & {} & {} & {} & {noise$^a$}\\ 
{} & {(GHz)~} & {(km.s$^{-1}$)~} & {(km.s$^{-1}$)~} & {(K)~} \\ 
\hline \\
{$^{12}$CO(2--1)} & {230.5379} & {0.84} & {157} & 0.1 \\
{$^{13}$CO(2--1)} & {220.3986} & {0.10} & {78} & 0.2 \\ 
{C$^{18}$O(2--1)}& {219.5603} & {0.10} & {78} & 0.3 \\ 
 \hline 
\end{tabular} \label{tab:vespa_tab}
\vspace*{-0.45ex}
\begin{list}{}{}
\item[$^a$]{ {\bf rms noise per spectral resolution element. } } 
\end{list} 
\end{table} 

We also mapped the NGC~2264-C region in the 450 $\mu$m  dust continuum 
at $\sim 10\arcsec $ resolution (HPBW) 
with the P-ArT\'eMiS bolometer array on the Atacama Pathfinder Experiment (APEX) telescope 
in November 2007. 
The resulting map is shown in Fig.~\ref{fig:cont}(b), and details about the P-ArT\'eMiS camera 
and the November 2007 run at APEX can be found in \citet{andre2008}.

\subsection{Data reduction} 
All data were reduced with the CLASS90 program, a new part of the GILDAS software package. Dedicated CLASS90 scripts were written to automatically reduce the data and produce a datacube for each of the three observed transitions (see Table \ref{tab:vespa_tab}).
\newline Linear (first-order) baselines were determined from velocities ranges where emission is no longer positively detected above the 3$\sigma$ noise limit anywhere in the map, and then subtracted from the spectra. 
We stress that our HERA observations resulted in very good baselines  (see example in Fig.~\ref{fig:bound_F4}).
We used 5 km.s$^{-1}$-wide velocity windows at either end of the $^{12}$CO(2--1) spectra 
to fit the substracted baselines, making us sensitive to any outflow emission present up to LSR velocities of 
V$_{sys} \pm$ 55 km.s$^{-1}$.
\newline Some velocity channels showed spikes due to band splitting of the VESPA autocorrelator. 
We had to flag them and replace their values by interpolating the closest two good channels. 
The worst case occurred around v$_{LSR}\sim +23$ km$/$s with four consecutive bad channels, 
leading to an interpolation over $\sim$ 1.7 km$/$s, which we consider as being negligible for the present study of high-velocity emission.
\newline Finally, in order to improve the signal-to-noise ratio, we also smoothed the initial velocity resolution by a factor of two for the $^{12}$CO(2--1) map, leading to a final spectral resolution of 0.84 km$/$s per channel.

\begin{figure*}[!ht]
\centering
\includegraphics[width= 0.8\linewidth,angle=-90]{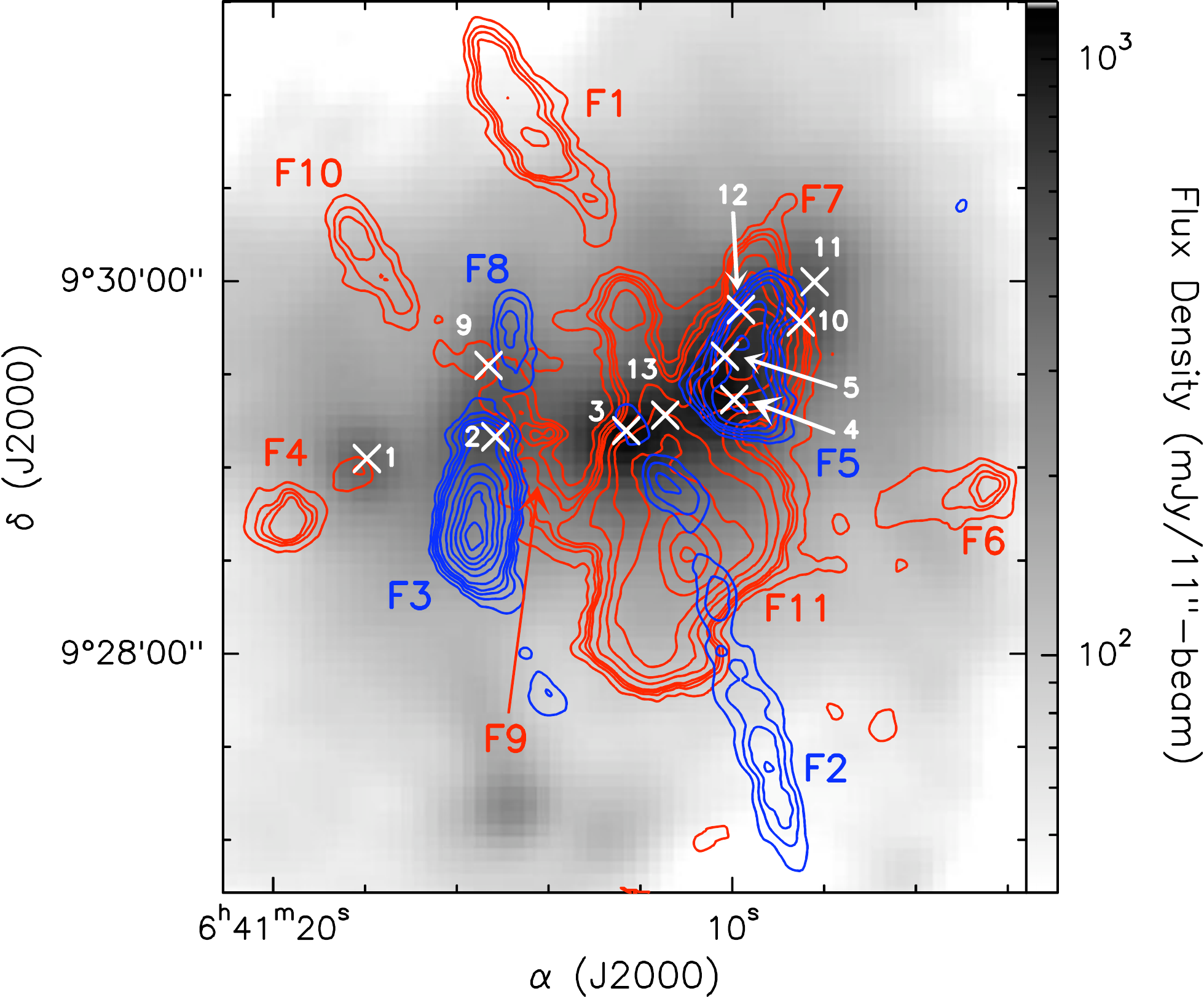}\\
\caption{$^{12}$CO(2--1) map of the NGC~2264-C protocluster. Blue contours show the levels of $^{12}$CO(2--1) intensity integrated between $-27$ km/s and 2 km/s in the blue-shifted part of the line, and go from 5 to 98 K.km.s$^{-1}$. Red contours are levels of intensity integrated between 13 km/s and 34 km/s in the red-shifted part of the line, and go 
from 5 to  110 K.km.s$^{-1}$. The eleven outflows discovered in the present survey are labeled by F1 to F11. The background greyscale image is the 1.2~mm dust continuum map obtained by \citet{peretto2006} with MAMBO on the IRAM-30m. White crosses and numbers 
refer to the millimeter continuum peaks identified by \citet{peretto2006,peretto2007} 
(see also Fig.~\ref{fig:cont}).}
\label{fig:allmap}
\end{figure*}

\section{Mapping results and analysis} 
 
We detected a total of eleven sub-regions or "lobes" exhibiting high-velocity emission in the $^{12}$CO(2--1) map (Fig. \ref{fig:allmap}). 
These eleven lobes are spatially distributed around the millimeter continuum cores identified by  \citet{peretto2006, peretto2007}, and four of these lobes can be directly associated with Class~0 - like objects, as discussed in \S~5.1 below.

\noindent In \S3.1 we explore the kinematical properties of these eleven outflow lobes, while \S3.2 discusses the optical depth 
 of the $^{12}$CO(2--1) transition in our map.

\subsection{Spatial extent of the outflow lobes and minimum associated velocity intervals.} 
In order to determine the spatial extent of each outflow, we need to find out which pixels exhibit high-velocity emission due to outflows. We know from previous work that the NGC~2264-C protocluster exhibits a complex pattern of cloud velocities: the mean LSR velocity of the cloud was found to be 7.5$\pm$0.2 km$/$s, but a pronounced velocity discontinuity was also found by \citet{peretto2006} near CMM3, and local systemic velocities can depart from this value by up to 0.7~km$/$s, depending on position in the protocluster. 
Likewise, the intrinsic $^{12}$CO(2--1) linewidth varies with position as column density and turbulent motions vary through the protocluster. 
Therefore, to discriminate emission due to high-velocity outflow motions from
emission due to an intrinsically wider  $^{12}$CO(2--1) line from the cloud itself, and thus identify those pixels with outflow emission, we adopted the following conservative method.

\begin{figure}[!h]
\begin{center}
  \subfigure[]{\includegraphics[width=0.5\linewidth,angle=-90]{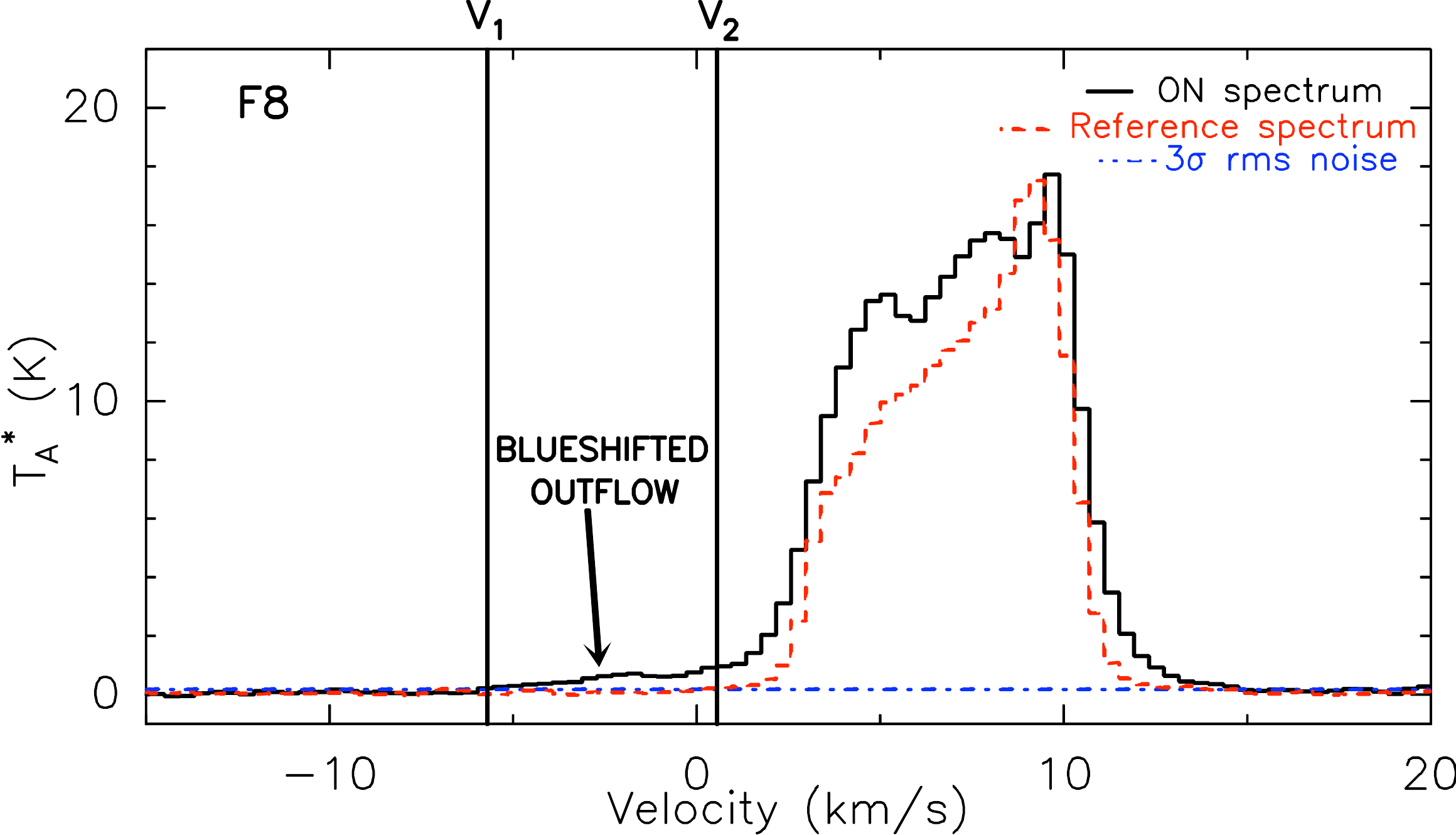}}
  \\
  \subfigure[]{\includegraphics[width=0.5\linewidth,angle=-90]{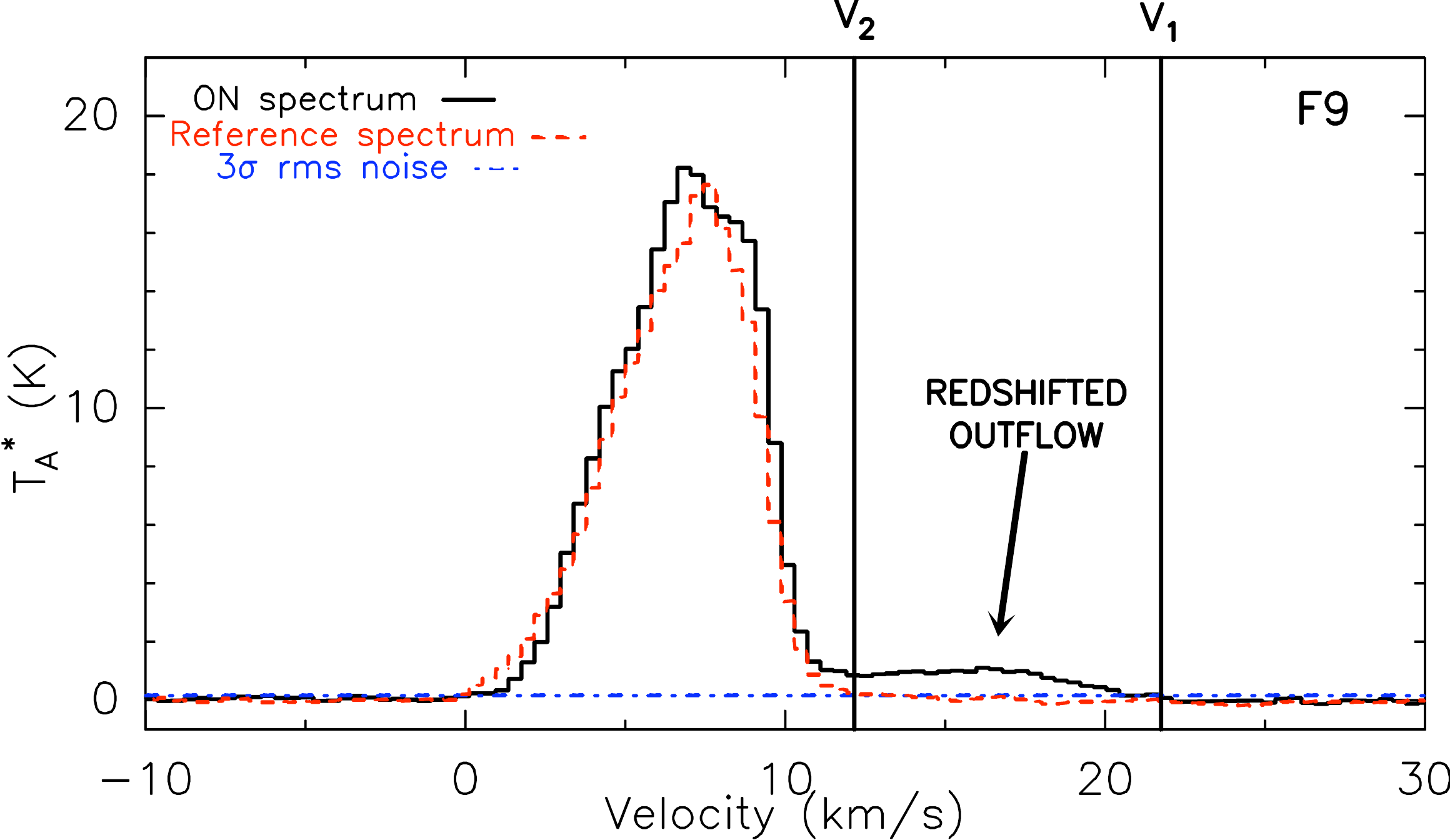}}
\caption{Examples of high-velocity $^{12}$CO(2--1) spectra showing emission at blue-shifted (a) and red-shifted (b) velocities for the candidate outflows 
F8 and F9, respectively. 
In both cases, the dashed spectrum shows a $^{12}$CO(2--1) reference spectrum obtained by averaging $\sim$100 spectra from the ambient cloud emission 
in the immediate vicinity of the outflow. 
Vertical solid lines highlight the velocity range over which emission is no longer detected from the ambient cloud, 
whereas blue-shifted or red-shifted high-velocity emission is still present in the spectrum (above the $3\sigma$ noise level shown as an horizontal dot-dashed line)} 
\label{fig:high_velo}
\end{center}
\end{figure}

We first constructed $^{12}$CO(2--1) reference spectra for the local ambient medium surrounding  the high-velocity lobes. 
In practice, we constructed each reference spectrum by averaging a large number of spectra (from 30 to 200 spectra depending on the subregion considered) located in the close neighborhood of each candidate outflow.
In this way, we obtained one reference spectrum per high-velocity lobe for outflows F3 to F11. 
We constructed two reference spectra for each of F1 and F2, so as to ensure that 
small variations in the systemic velocity occurring between different sub-regions of these extended lobes be properly accounted for.
In a second step, we compared each reference spectrum with the corresponding high-velocity lobe line profiles.
Each pixel which showed emission stronger than three times the typical noise level, $\sigma_{\rm{spec}}$,  in the $^{12}$CO(2--1) outflow spectrum in three consecutive channels 
(corresponding to a 2.5 km$/$s-wide window in velocity), at velocities where emission was no longer detected in the reference spectrum,  was assigned to the candidate outflow under consideration. 

\begin{figure}[!h]
\begin{center}
    \includegraphics[width= 0.8\columnwidth,angle=0]{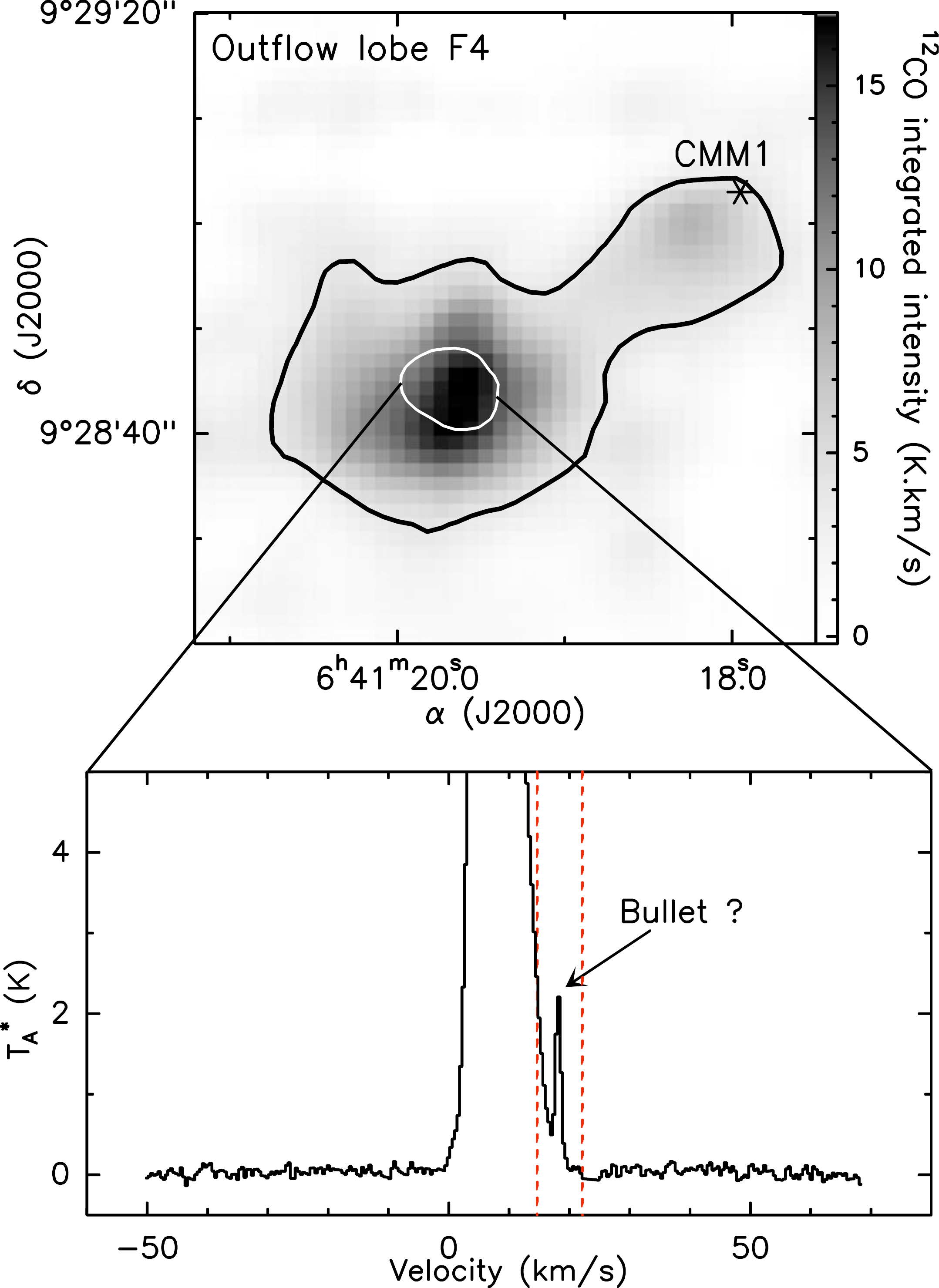}
    \caption{Spatial extent of the red-shifted outflow lobe F4.
    The upper panel shows the high-velocity $^{12}$CO(2--1) emission, integrated from $+14.3~{\rm km.s}^{-1}$ to $+23~{\rm km.s}^{-1}$ 
    (velocity interval marked by dashed lines in the lower panel).
    In this case, the {\it reference} spectrum associated to F4 does not show any significant emission 
    at LSR velocities larger than $14.3~{\rm km.s}^{-1}$.
     The 3$\sigma_{spec}$ level mentioned in the text corresponds to the thick black contour. 
     All pixels showing integrated emission above this value 
     were therefore assigned to outflow F4. 
A typical $^{12}$CO(2--1) spectrum observed near the peak of outflow emission is shown in the lower panel.
    Note the quality of the baseline and the detached high-velocity feature at v$_{LSR} \sim +18$~km.s$^{-1}$.
    The area over which such a ``bullet-like'' emission feature (see \S ~5.1) is seen in F4 is marked by the white contour in the upper panel.}
      \label{fig:bound_F4}
     \end{center}
\end{figure}

%
Such cases are illustrated in Fig.~\ref{fig:high_velo}, using spectra from the two outflows F8 and F9, for which obvious high-velocity wings are observed, at blueshifted and redshifted 
velocities, respectively. The 
vertical solid lines mark the velocity range over which emission is only due to outflow material: in this velocity range no emission is detected in the reference spectrum, 
while emission is positively detected in the spectra from the candidate outflow. 
Here, we ignore any additional outflow emission at lower velocities where emission is still seen in the reference spectrum. 
Distinguishing between low-velocity outflow emission and turbulent broadening of the line spectrum from the ambient cloud is more uncertain.
We will try to account for the low-velocity outflow emission when we estimate outflow parameters in \S ~4 below.
For the purpose of determining the spatial extents of the outflows, however, only those  pixels satisfying the selection criteria described above were included.

Following this procedure for all pixels in the map, we were able to determine the spatial extent of 
each outflow lobe (see Col.~4 of Table~\ref{tab:small_tab} for the values, and Fig.~\ref{fig:bound_F4} for an example). 
The velocity range over which high-velocity emission was detected from the outflow lobe 
but no emission was seen in the reference spectrum will be adopted as the 
main LSR velocities interval to be considered in further calculations.
In the following, $\vert \rm{V}_{1}\vert$ will denote the extreme LSR velocity observed toward the outflow 
(beyond which T$_{outflow} < 3\sigma_{spec}$), 
and $\vert \rm{V}_{2}\vert$ will denote the minimum LSR velocity at which unambiguous outflow emission 
is detected (beyond which T$_{ref} < 3\sigma_{spec}$ while T$_{outflow} > 3\sigma_{spec}$).
As can be seen in Table \ref{tab:small_tab}, the velocity ranges spanned by the observed high 
velocity emission vary from outflow to outflow, 
and $^{12}$CO(2--1) emission is detected  at very high LSR velocities in some cases 
(up to  $-27$~km/s  in the case of the blue-shifted lobe F2).

\begin{table*}[!ht]
\begin {center}
\centering\par\caption{Observed properties of the $^{12}$CO(2--1) outflows detected in the map of Fig.~2}
\begin{tabular}{cccccc}
\hline
\hline
{OUTFLOW} & {Local systemic} $^{(1)}$& \multicolumn{2}{c}{Outflow main $^{(2)}$} & {Area} $^{(3)}$& {Length} $^{(4)}$\\
{LOBE}  & {velocity $\rm{V_{sys}}$} & \multicolumn{2}{c}{LSR velocities interval} & {} & {} \\
{}   & {(km.s$^{-1}$)} & \multicolumn{2}{c}{(km.s$^{-1}$)} &  {(arcsec$^{2}$)} & {(arcsec)} \\
       \hfill & \hfill & {~~~~~V$_{1}$~~~~} & {V$_{2}$} & \hfill & \hfill \\
\hline
	{F1 (Red)} & 7.5 & 28.5 & 13 & 2780 & 127 \\
	{F2 (Blue)}& 7.6 & $-$27 & 1.1 & 3060 & 184\\
	{F3 (Blue)}& 7.4 & $-$26.6 & 1.4  & 1935  & 70 \\
{F4 (Red)}& 7.7 & 23 & 14.3 & 900  & 53 \\
{F5 (Blue)} & 7.3 & $-$12 & 0.3 & 1080  & 50 \\
{F6 (Red)}& 7.1 & 28.1 & 13.1 & 1360  & 56 \\
{F7 (Red)}& 7.2 & 33.2 & 14.7 & 2230   & 72 \\
{ F8 (Blue)}&  7.2 & $-$23 & 1.7 & 590 & 38 \\
{ F9 (Red)} & 7.2 & 30.2 & 13.7 & 800  & 53 \\
{ F10 (Red)}& 7.2 & 32.2 & 13.7 & 610  & 50 \\
{ F11 (Red)} & 7.7 & 33 & 15.2 & 4180  & 90 \\
\hline

\end{tabular}
 \vspace*{-0.45ex}
\begin{list}{}{}
\item[$^{(1)}$]{Systemic LSR velocity of the ambient material around each outflow.} 
\item[$^{(2)}$]{Velocity interval [V$_{1}$ ; V$_{2}$] over which each outflow emission is detected but no significant emission is seen in the reference spectrum.} 
\item[$^{(3)}$]{Surface area of each outflow lobe.}
\item[$^{(4)}$]{Major size of each outflow lobe.}
\end{list}
\label{tab:small_tab}
\end {center} 
\end{table*}

\subsection{Optical depth of the $^{12}$CO(2--1) emission}

In order to derive the masses and moments for the eleven candidate outflows, we need to assess 
the optical depth of the CO emission. 
We used our observations in three isotopes (see Table \ref{tab:vespa_tab}) to estimate the optical depth of 
the CO(2--1) transition in our data. 

We first used our C$^{18}$O(2--1) data to check that the $^{13}$CO(2--1) emission observed away from line center was optically thin 
over the entire mapped area and estimated a maximum optical depth $\tau \left( ^{13} \rm{CO}\right) \sim 0.75$ at the systemic velocity 
of the ambient cloud.

\begin{figure}[!h]
\centering
\includegraphics[width= 0.9\columnwidth,angle=0]{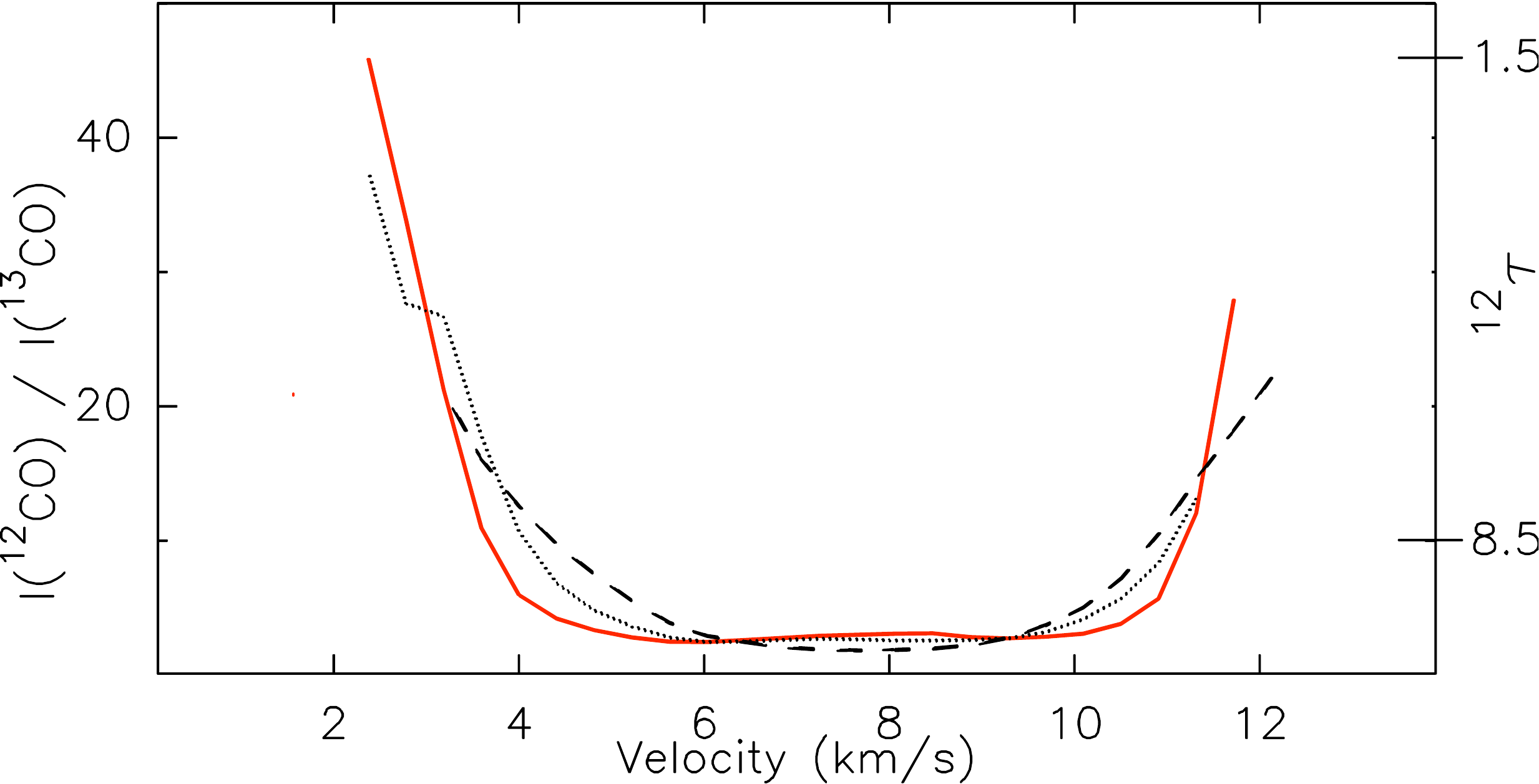}\\
\caption{$^{12}$CO(2--1) to $^{13}$CO(2--1) intensity ratio calculated over the widest velocity range for which the $^{13}$CO(2--1) intensity is larger than the 3$\sigma$ noise limit. 
Outflow F4 is represented as a solid line, outflow F7 as a dashed line, and outflow F3 as a dotted line. The vertical axis on the right shows the corresponding values of 
the $^{12}$CO(2--1) optical depth, $^{12}\tau$(v).}
\label{fig:lineratios}
\end{figure}

We then derived the $^{12}$CO(2--1) optical depth in the high-velocity gas 
from the observed $^{12}$CO/$^{13}$CO line intensity ratio (cf. \citealt{goldsmith1984}). 
We measured this ratio at eleven positions, chosen to coincide with the positions of the candidate 
outflows in the $^{12}$CO  map.
Assuming that the excitation temperatures of the $^{12}$CO and $^{13}$CO lines are equal, and 
knowing that the $^{13}$CO emission is optically thin, the $^{12}$CO optical depth $^{12}\tau$(\rm{v}) at any 
given velocity v is related to the $^{12}$CO/$^{13}$CO intensity ratio by: 
\begin {equation}
\frac{I\left(^{12}CO\right)_{\rm{v}}}{I\left(^{13}CO\right)_{\rm{v}}} =\frac{1-\exp{^{- ^{12}\tau(\rm{v})}}}{^{13}\tau(\rm{v})}= \frac{1-\exp{^{- ^{12}\tau(\rm{v})}}}{^{12}\tau(\rm{v})} \cdot X_{iso}, 
\end {equation}

\noindent where  X$_{iso}$ is the [$^{12}$CO]/[$^{13}$CO] abundance ratio, here assumed to have the terrestrial value of 89 (\citealt{wilson1992}). 
The signal-to-noise ratio in the $^{13}$CO(2--1) spectra was not adequate to derive $^{12}\tau_{\rm{v}}$ over the full velocity extent of the line wings seen 
in $^{12}$CO(2--1). 
We calculated the isotopic ratio up to velocities where the $^{13}$CO(2--1) emission became lower than the 3$\sigma$ noise limit, on either side of the line centroid for the eleven outflows.
Figure~\ref{fig:lineratios}  shows the $^{12}$CO(2--1)  to $^{13}$CO(2--1) ratio for three different outflows (outflow F4 as a solid line, outflow F7 as a dashed line, 
and outflow F3 as a dotted line). 

\noindent The  $^{12}$CO(2--1)/$^{13}$CO(2--1) intensity ratio is expected to increase, and $^{12}\tau_{\rm{v}}$ to decrease, at larger velocity offsets from the line centroid, 
but the $^{13}$CO(2--1) emission becomes undetectable at these high velocities. 
At the highest velocities showing detectable $^{13}$CO(2--1) emission, a maximum $^{12}$CO(2--1)/$^{13}$CO(2--1) intensity ratio of $\sim$~45 is measured 
for both blueshifted and redshifted gas (cf. Fig.~\ref{fig:lineratios}), which corresponds to $^{12}\tau_{wings} \sim 1.5$.
Therefore,  $^{12}\tau$ is much lower than 1.5 at LSR velocities larger than 12~km/s or smaller than 2~km/s. 
As the velocity intervals derived in \S ~3.1.2 for the candidate outflows lie outside the [2 : 12]~km/s range, we conclude 
that the $^{12}$CO(2--1) outflow emission is optically thin at both blue-shifted and red-shifted velocities.

Note, however, that we will apply an optical depth correction factor when we derive the amount of mass entrained by the outflows at 
low velocities (i.e., inside the [2 : 12]~km/s range)  in \S ~4.1.2 below.
For this purpose, we will use the values of $^{12}\tau(\rm{v})$ estimated at LSR velocities between $+2$  and $+12$~km/s (see Figure~\ref{fig:lineratios} for examples).


\section{Momentum injection rate due to outflows in NGC~2264-C}

Our $^{12}$CO(2--1) map revealed the presence of eleven outflows  in the NGC~2264-C protocluster.
To quantify the effective feedback that these eleven outflows produce on the protocluster, we now estimate the masses of 
outflow entrained material, as well as the outflow dynamical parameters.
\subsection{Outflow mass estimates}

\subsubsection{Method of derivation}
Assuming local thermal equilibrium (LTE), 
the total gas mass of an outflow can be obtained from the observed $^{12}$CO(2--1) spectra as follows (see, e.g.,  \citet{scoville1986}):
\begin {equation}
M_{\rm{flow}} = 5.3 \times10^{-8} \times \frac{T_{ex}+0.93}{e^{\frac{-16.77}{T_{ex}}}}
\times d^{2} \times
\int_{\rm{V_{a}}}^{\rm{V_{b}}}\int_{A} I(\rm{v})~\it{d}\rm{v} . \it{d}\rm{A}
~~(M_\odot), 
\end {equation}
where $T_{ex} $ is the CO excitation temperature (in Kelvin), and $d$ the distance to the source expressed in kpc. 
$I(\rm{v})$ is related to the observed antenna temperature $T_{A}^{*}(\rm{v})$ (expressed in Kelvin) by 
$I(\rm{v})=\left(\frac{\tau_{\rm{v}}}{1-e^{-\tau_{\rm{v}}}}\right)\times T_{A}^{*}(\rm{v})$, 
and $\int_{\rm{V_{a}}}^{\rm{V_{b}}}\int_{A} I(\rm{v})~\it{d}\rm{v} . \it{d}\rm{A}$ (with $I(\rm{v})$ in K, $d$v in km.s$^{-1}$, and $d$A in arcsec $^{2}$) 
is the $^{12}$CO(2--1) intensity integrated over the surface area A,  and over the velocity interval 
[V$_{\rm{a}}$ : V$_{\rm{b}}$] of the outflow.
In deriving this formula, we adopted a standard CO to H$_{2}$ abundance ratio, $[\frac{CO}{H_{2}}] = 10^{-4}$ 
(Frerking, Langer, \& Wilson 1982).
\subsubsection{Masses of outflow-entrained gas}

\begin{figure}[ht]
\centering
\includegraphics[width= 0.7\columnwidth,angle=-90]{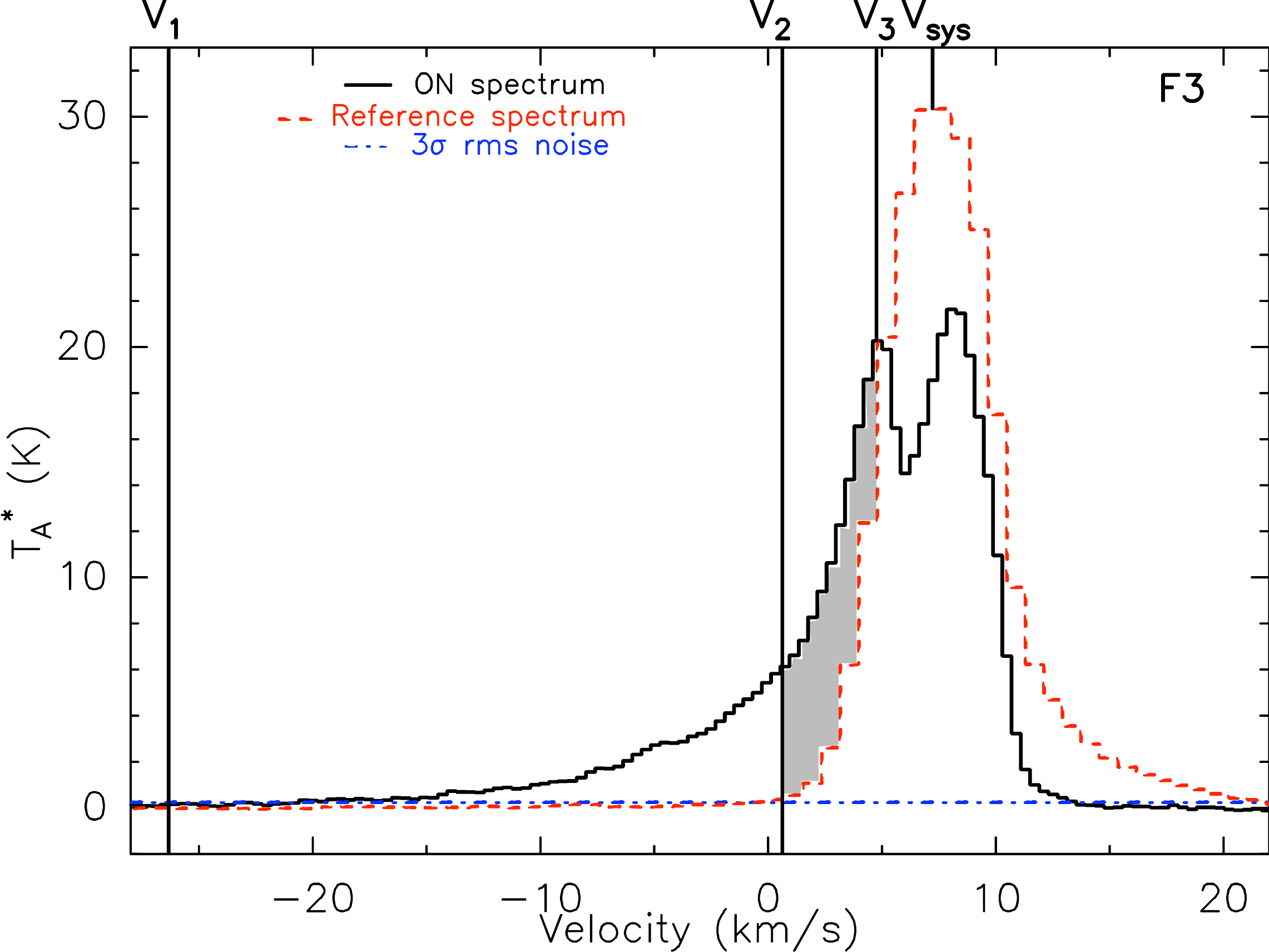}
\caption{Illustration of the velocity ranges used for the calculation of minimum and maximum masses of the outflows. 
The dashed spectrum shows the $^{12}$CO(2--1) reference spectrum from the ambient cloud.  
The solid spectrum corresponds to the candidate outflow F3. Vertical solid lines show the velocity ranges used in the computation of the outflow masses. 
[V$_{1}$ ; V$_{2}$] represents the interval used for the calculation of the minimum outflow mass (see \S4.1.2 for details). [V$_{2}$ ; V$_{3}$] is the additional 
velocity interval used to compute M$_{add}$ by integrating the excess $^{12}$CO(2--1) emission (area highlighted in grey in this case).}
\label{fig:mass_estim_F3}
\end{figure}

~~~~ We first computed the {\it minimum} mass of gas associated with each outflow, assuming 
that the whole outflow emission is included in the main LSR velocity range [V$_{1}$ : V$_{2}$] derived 
in \S~3.1.2 (Col. 3 of Table~\ref{tab:small_tab}). 
In this first estimate, we thus did not take into account any additional outflow emission seen at 
lower absolute velocity offsets from line center, where emission is positively detected 
in the reference spectrum (see Fig.~\ref{fig:mass_estim_F3}). 
We showed in \S~3.2 that the $^{12}$CO(2--1) emission is optically thin everywhere outside 
the velocity range [2 : 12]~km/s in our map. 
As the minimum velocity intervals considered in this first estimate all lie outside the [2 : 12]~km/s range, 
the minimum masses of outflowing gas can thus be computed as:
\begin{equation}
M_{\rm{min}} = 1.6 \times 10^{-6} \times \int_{\rm{V_{1}}}^{\rm{V_{2}}}\int_{A} T_{A}^{*}(\rm{v})~{\it{d}}v . {\it{d}}A
 ~~(M_\odot)
 \end{equation}
\noindent In the above equation, $T_{A}^{*}(\rm{v})$ is expressed in Kelvin, {\it{d}}v is a velocity bin (in km.s$^{-1}$), while {\it{d}}A corresponds to the surface element (in arcsec$^{2}$).
\newline\noindent The main source of uncertainty on the derived outflow-entrained masses is the uncertain 
value of the CO excitation temperature. 
For optically thin CO emission, the excitation temperature is expected to be close 
to the gas kinetic temperature due to the low dipole moment of the CO molecule. 
Nevertheless, large temperatures (up to 80~K), due to gas heating by outflow shocks 
(\citealt{umemoto1992,bachiller1993,bachiller2001}) are sometimes found in CO outflows.
As can be seen from Eq.~(2),  the derived outflow mass is almost proportional to the assumed excitation temperature. 
We assumed an excitation temperature of 20~K in our calculations, which is representative of typical CO ouflows. 
The uncertainty in the CO excitation temperature introduces a factor of $\sim 2$ uncertainty on our 
estimates of the masses, momenta,  and momentum fluxes.
%
%
%
%
%
\newline We calculated the value of $\int_{\rm{V_{1}}}^{\rm{V_{2}}}\int_{A} T_{A}^{*}(v)~dv . dA$ directly from our 
data by integrating the intensity of the $^{12}$CO(2--1) emission, both spatially over the outflow extent, 
and kinematically over the main LSR velocities range [V$_{1}$ ; V$_{2}$] of each outflow. The outflow extents and main LSR 
velocities ranges adopted are given in Table~\ref{tab:small_tab}. \\

\noindent The minimum mass of gas associated with each outflow is given in the fifth column 
 (first value) of Table~\ref{tab:big_tab}.
The minimum total mass of gas entrained by the network of eleven outflows in NGC~2264-C is 
estimated to be 0.3$~\pm~$0.15 M$_\odot$.

In order to estimate the {\it maximum} mass of gas entrained by the eleven outflows observed, two calculations were made.
\newline First, for each outflow lobe, we assumed that all of the blue-shifted or red-shifted emission such as T$_{\rm{outflow}} >$T$_{\rm{ref}} + 3\sigma_{\rm{spec}}$ 
was due to the corresponding outflow (see Fig.~\ref{fig:mass_estim_F3}).
This hypothesis may overestimate outflow masses by including low-velocity motions due to local turbulent broadening of the ambient cloud emission.
The corresponding values therefore provide upper limits to the masses of outflow-entrained gas. 
We also recall that,  at low velocities, the $^{12}$CO(2--1) emission is no longer optically thin (see Fig.~\ref{fig:lineratios}). 
Therefore, we used the values of optical depth derived previously to correct for optical depth effects on our mass estimates.
The additional outflow emission at low velocities (see Col.~3 of the online Table~\ref{tab:addmass_tab} for numerical values) can be expressed as:
\begin{equation}
M_{\rm{add}} = 1.6 \times 10^{-6}\times
\int_{\rm{V_{2}}}^{\rm{V_{3}}}\int_{A}\frac{\tau_{\rm{v}}}{1-e^{-\tau_{\rm{v}}}}~[T_{A}^{*}(\rm{v})\vert_{flow} - T_{A}^{*}(\rm{v})\vert_{ref}] d\rm{v} . d\rm{A}   ~~(M_\odot), 
\end{equation}
where  $\vert{V_{2}}\vert$ is the velocity at which $T_{A}^{*}(\rm{v})\vert_{ref} < 3\sigma_{ref}$
\newline and $\vert{V_{3}}\vert$ the velocity at which $T_{A}^{*}(\rm{v})\vert_{flow}$ = $T_{A}^{*}(\rm{v})\vert_{ref} - 3\sigma_{spec}$ \\
(if $T_{A}^{*}(\rm{V_{sys}})\vert_{flow} > T_{A}^{*}(\rm{V_{sys}})\vert_{ref}$ then $\vert{\rm{V_{3}}}\vert = \rm{V_{sys}}$).
\newline\noindent As a second step, we further took into account the material entrained by outflows down to velocities comparable to the systemic velocity of the surrounding medium.
We first extracted a characteristic mean spectrum per outflow, and fitted a power law to this mean spectrum, 
$T_{\rm{outflow}} = \beta \times (\rm{v}-\rm{V}_{sys})^{\alpha}$, in the velocity range $\rm{V}_{1} < \rm{v}< \rm{V}_{3}$. 
We then extrapolated the fitted profile for velocities $\rm{V}_{3} < \rm{v} <  \rm{V}_{sys}$, 
in order to estimate the amount of mass entrained at very low velocities and which remains hidden in the main body of the $^{12}$CO(2--1)  line: 
\begin{equation}
M_{\rm{low}} = 1.6 \times 10^{-6} \times
\int_{\rm{V_{3}}}^{\rm{V_{sys}}}\int_{A}\frac{\tau_{\rm{v}}}{1-e^{-\tau_{\rm{v}}}}
~[\beta \times (\rm{v}-\rm{V}_{sys})^{\alpha}] .\rm{dv} \times \rm{A}
~~~~(M_\odot), 
\end{equation}
where $\rm{\rm{V}}_{\rm{sys}}$ is the systemic velocity associated with the outflow (see Col.~2 of Table \ref{tab:small_tab}).
 \newline The power-law indices of the fit can be found in the Col.~4 of the online Table~\ref{tab:addmass_tab}. 
The additional mass of outflow-entrained gas at velocities between $\rm{V}_{3}$ and the systemic velocity of the cloud $\rm{V}_{sys}$ is given in 
Col.~5 of the online Table~\ref{tab:addmass_tab}.
We obtained satisfactory fits for all eleven outflow lobes, except outflow F4 which exhibits a detached high-velocity emission feature (see Fig.~\ref{fig:bound_F4}). 
The best fits lead to a fraction of invisible, low-velocity outflow-entrained mass $M_{\rm{low}}$/$M_{\rm{max}} \sim$ $10\%$ -- $45\%$. 
Since this additional mass is moving at very low velocities (1 to 3 km.s$^{-1}$ away from the systemic velocity), its contribution to the 
outflow dynamical parameters will nevertheless be quite small (see \S~4.2 below).

Finally, the maximum masses of outflow-entrained gas $M_{\rm{max}}$ were computed by adding the masses $M_{\rm{add}}$~+~$M_{{\rm{low}}}$ 
calculated in the velocity interval [V$_{2}$ ; V$_{\rm{sys}}$] to the minimum masses $M_{\rm{min}}$ derived earlier.
These maximum masses are reported in Col.~5 of Table~\ref{tab:big_tab} (second value). 
We stress that these masses represent the maximum possible values of the outflow masses, as the hypothesis made in their derivation 
led us to include all of the $^{12}$CO emission potentially due to outflows.
The overall uncertainty on each of the outflow mass components ($M_{\rm{min}}$, $M_{\rm{add}}$, $M_{\rm{low}}$) is about a factor of two.

%
%
\onltab{3}{
\begin{table*}
\begin {center}
\centering \par \caption{Estimates of the masses of outflow-entrained material at low velocities} 
\begin{tabular}{ccc|cc}
\hline
\hline
{OUTFLOW} & {Additional velocity}  & {Additional mass} & {Index of} & {Additional} \\
{LOBE}  & {interval [V$_{2}$ ; V$_{3}$]} & {$M_{\rm{add}}$ in [V$_{2}$ ; V$_{3}$]} & {power-law fit, $\alpha$}  & {mass $M_{\rm{low}}$ in [V$_{3}$ ; V$_{\rm sys}$]}\\
       & {(km.s$^{-1}$)} & {($10^{-2}$~M$_\odot $)} &  {} & {($10^{-2}$~M$_\odot $)} \\
       \hfill & $^{(1)}$ & $^{(2)}$ & $^{(3)}$ & $^{(4)}$ \\
\hline
{F1 (Red)} & [13 ; 10] & 7 $\pm$ 2& $-$2.7 $\pm$ 0.6 & 25 \\
{F2 (Blue)}& [1.1 ; 4.3] & 1.9 $\pm$ 0.5 & $-$2.5 $\pm$ 0.3 & 32 \\
{F3 (Blue)}& [1.4 ; 4.9] & 6 $\pm$ 3 & $-$2.3 $\pm$ 0.3 & 24 \\
{F4 (Red)}& [13 ; 10.2] & 2.3 $\pm$ 1.5 & - & - \\
{F5 (Blue)} & [0.3 ; 5] & 6 $\pm$ 2.8 & $-$3.1 $\pm$ 0.9 & 31 \\
{F6 (Red)}& [13.1 ; 9.9] & 5.7 $\pm$ 3.1 & $-$3.7 $\pm$ 0.8 & 26 \\
{F7 (Red)}& [14.7 ; 10] & 15 $\pm$ 7 & $-$2.8 $\pm$ 0.7 & 90 \\
{ F8 (Blue)}& [1.7 ; 4.4] & 0.37 $\pm$ 0.15 & $-$2.4 $\pm$ 0.4 & 7  \\
{ F9 (Red)} & [13.7 ; 9.6] & 1.2 $\pm$ 0.7 & $-$2.4 $\pm$ 0.5  & 6 \\
{ F10 (Red)}& [13.7 ; 9.7] & 1 $\pm$ 0.55 & $-$2.5 $\pm$ 0.3 & 7 \\
{ F11 (Red)} & [15.2 ; 9.5] & 30 $\pm$ 18 & $-$2.0 $\pm$ 0.9 & 52 \\
\hline
\end{tabular}
\label{tab:addmass_tab}
\vspace*{-0.45ex}
\begin{list}{}{}
\item[$^{(1)}$]{Velocity range [V$_{2}$ ; V$_{3}$] in which $T_{A}^{*}(v)\vert_{ref} > 3\sigma_{ref}$ and $T_{A}^{*}(v)\vert_{outflow} > T_{A}^{*}(v)\vert_{ref} - 3\sigma_{ref}$.}
\item[$^{(2)}$]{Mass obtained by integration of [$T_{A}^{*}(v)\vert_{outflow} - T_{A}^{*}(v)\vert_{ref}$] over [V$_{2}$ ; V$_{3}$].}
\item[$^{(3)}$]{Power-law index of the best fit on line wings observed in [V$_{1}$ ; V$_{3}$].} 
\item[$^{(4)}$]{Estimate of the  mass which remains hidden in the main body of the line [V$_{3}$ ; V$_{sys}$].}
\end{list}
\end {center} 
\end{table*}
}
%
%

\subsection{Outflow dynamical parameters}
~~~ Using the mass estimates given in Col.~5 of Table~\ref{tab:big_tab}, 
we first computed the momentum associated with each outflow:
\begin{equation}
P_{\rm{flow}} = M_{\rm{flow}}\times \rm{V}_{\rm{char}}
\end{equation}
where $\rm{V}_{\rm{char}}$ is the characteristic velocity of the outflow lobe, 
computed as the intensity-weighted mean absolute velocity of the outflow over the relevant velocity range : 
\begin{center}
$\rm{V}_{\rm{char}}$ =  {\large{$\frac{\rm{V}_{\rm{char}}^{obs}}{\cos(i)}$}} = {\large{$\frac{1}{N_{pix} \times \cos(i)}~\sum \limits_{pixels}$}}
{\large{$\frac{\sum \limits_{\Delta \rm{v}}~\rm{T}_{\rm{A}}^{\star}(\rm{v}) \vert \rm{v}-\rm{V}_{\rm{sys}}\vert}
{\sum \limits_{\Delta \rm{v}}~\rm{T}_{\rm{A}}^{\star}(\rm{v})}$}}.
\end{center}

The inclination angle $i$ between the flow axis and the 
line of sight (l.o.s.) has to be taken into account when computing the true characteristic 
velocity $\rm{V}_{\rm{char}}$, since it is related to the observed characteristic velocity 
(projected on the l.o.s.) $\rm{V}_{\rm{char}}^{\rm{obs}}$ (see Col.~2. of Table~\ref{tab:big_tab}), 
by $\rm{V}_{\rm{char}}$ = $\rm{V}_{\rm{char}}^{\rm{obs}}~/ \cos(i)$.
\\We derived the outflow momentum in each of the three velocity regimes  [V$_{1}$ ; V$_{2}$],  [V$_{2}$ ; V$_{3}$],  [V$_{3}$ ; V$_{\rm{sys}}$] considered 
for the calculations of the outflow-entrained mass (cf. \S ~4.1.2 and Fig.~\ref{fig:mass_estim_F3}). 
The total outflow momentum was calculated as the sum of the three momenta estimated 
in the three velocity regimes.

The outflow momentum flux was obtained by dividing the momentum by 
the estimated dynamical timescale  {\large{$t_{\rm{dyn}}$}} of the outflow: 
\begin{equation}
F_{\rm{flow}} = P_{\rm{flow}} / t_{\rm{dyn}}.
\end{equation}
The dynamical timescale, {\large{$t_{\rm{dyn}}= l_{\rm{flow}} / \rm{V}_{\rm{max}}$}}, 
also depends on the inclination angle $i$.  Indeed, $i$ affects both 
the estimate of the outflow length ({\large{$l_{\rm{flow}}= l~/ \sin(i)$}} 
where $l$ is the observed projected length on the sky), 
and the velocity estimates since the highest intrinsic flow velocity is 
V{\large{$_{\rm{max}}$}}={\large{$\rm{V}_{\rm{max}}^{\rm{obs}}$}} / $\cos(i)$, 
where $\rm{V}_{\rm{max}}^{\rm{obs}}$ is the highest observed velocity 
(corresponding to V$_{1}$ in \S3.1). 
This results in an inclination correction factor {\large{$\frac{\cos(i)}{\sin(i)}$}} for the dynamical timescale. 
Given the inclination correction on the momentum, the global inclination 
correction factor for the momentum flux is thus 
{\large{$f(i)=\frac{\sin(i)}{\cos^{2}(i)}$}} (cf. \citealt{bontemps1996}). 

%
%
%
%
\begin{table*}[!ht]
\begin {center}
\centering \par \caption{Derived parameters of the eleven $^{12}$CO(2--1) outflows mapped in NGC~2264-C.}
\begin{tabular}{ccccccc}
\hline
\hline
 {OUTFLOW}  & Characteristic velocity & Length & Dynamical time & Mass & Momentum &  Momentum flux \\
 {LOBE} & {V$_{\rm{char}}^{\rm{obs}}$} & {\small $i=90^{\circ}$ - $i=57.3^{\circ}$} & { \small $i=90^{\circ}$ - $i=57.3^{\circ}$} & {} & {} & {}\\
       		& {\normalsize (km.s$^{-1}$)} & {\normalsize (pc)} & {\normalsize ($10^{3}$ yr)} 
       		& {\normalsize($10^{-2}$ M$_\odot $)} & {\normalsize(M$_\odot$.km.s$^{-1}$)}
       		& {\normalsize($10^{-5}$ M$_\odot$.km.s$^{-1}$.yr$^{-1}$)}\\
		 \hfill & $^{(1)}$ & $^{(2)}$ & $^{(3)}$ & $^{(4)}$ & $^{(5)}$ & $^{(6)}$ \\ 
\hline
\vspace{-2 mm}
\hfill & \hfill & \hfill & \hfill & \hfill & \hfill & \hfill \\
{F1 (Red)} & $+$9.5 & 0.49 - 0.58 & 24.5 - 15.9 & 3.7 - 35.7 & 0.58 - 1.23 & 2.4 - 10.2 \\
{F2 (Blue)}& $-$13 & 0.71 -  0.84 & 20.0 - 13.0 & 2.9 - 36.8 & 0.37 - 0.97 & 1.8 - 6.8 \\
{F3 (Blue)}& $-$11.5 & 0.27 -  0.32 &  7.7 - 4.9 & 4.2 - 34.2 & 0.49 - 1.07 & 6.3 - 28.3 \\
{F4 (Red)}& $+$7.6 & 0.2 -  0.24 & 12.8 - 8.2 & 0.8 - 3.1 & 0.06 - 0.14 & 0.5 - 3.2 \\
{F5 (Blue)} & $-$12 & 0.18 -  0.21 & 9.1 - 5.8 & 0.9 - 37.9 & 0.12 - 0.74 & 1.3 - 11.5 \\
{F6 (Red)}& $+$6.6 & 0.2 - 0.24 & 9.3 - 5.9 & 1.6 - 33.3 & 0.1 - 0.71 & 1.1 - 10.5 \\
{F7 (Red)}& $+$10.3 & 0.28 - 0.33 & 10.6 - 6.8 & 6.0 - 111 & 0.62 - 2.56 & 5.8 - 35.3 \\
{F8 (Blue)}& $+$10.9 & 0.14 - 0.16 & 4.6 - 2.9 & 0.5 - 7.8 & 0.05 - 0.16 & 1.2 - 4.4 \\
{F9 (Red)} & $+$11 & 0.2 - 0.24 & 8.8 - 5.6 & 0.6 - 6.8 & 0.06 - 0.18 & 0.7 - 3.4 \\
{F10 (Red)} & $+$11 & 0.18 - 0.21 & 7.9 - 5.0 & 0.5 - 7.4 & 0.05 - 0.18 & 0.7 - 3.4 \\
{F11 (Red)} & $+$10.8 & 0.35 - 0.41 & 13.5 - 8.64 & 11.2 - 93.2 & 1.21 - 3.12 & 8.9 - 55.3  \\
\hline
\hline
\hfill & \hfill & \hfill & \hfill & \hfill & \hfill & \hfill \\
{L1551} $^{(7)}$& - & 0.65 - 0.77 & 48 - 31 & 530 - 670 & 18 - 23 & 47 - 136  \\
{L1157 (i$\sim$81$^{\circ}$)} $^{(7)}$& - & 0.4 & 2.5 &  62  & 31 & 1100 \\
\hline
\end{tabular}
\vspace*{-0.45ex}
\begin{list}{}{}
\item[$^{(1)}$]{Observed characteristic offset velocity of the outflow lobe (see details in \S4.2), computed over the [V$_{1}$~;~V$_{\rm{sys}}$] velocity range}
\item[$^{(2)}$]{Length of each outflow lobe.}
\item[$^{(3)}$]{Dynamical timescale of each outflow.}
\item[$^{(4)}$]{Minimum and maximum masses of outflow-entrained gas, computed as explained in \S ~4.1.}
\item[$^{(5)}$]{Minimum and maximum momenta, computed as explained in \S ~4.2.}
\item[$^{(6)}$]{Range of momentum flux derived for each outflow (see \S ~4.2.} 
\item[$^{(7)}$]{The last two rows give, for the sake of comparison, the values computed for the two bipolar outflows L1551 and L1157, used by \citet{nakamura2007}.}
\end{list}
\label{tab:big_tab}
\end {center} 
\end{table*}
%
%
%
%

The results of these calculations are given in Table~\ref{tab:big_tab}.
The length of each outflow is reported in Col.~3, where the first quoted value is the 
observed projected length of the outflow and 
the second value is corrected for the mean inclination angle i$=57.3^{\circ}$.
Column~4 gives the dynamical timescale of each outflow. The lower value is the apparent dynamical timescale (with no correction for inclination), 
while the upper value is corrected for the mean inclination angle i$=57.3^{\circ}$.
The outflow momentum and momentum flux values are given in Col.~6 and Col.~7.
The minimum quoted values were derived using the minimum mass estimates and 
ignoring inclination effects, i.e. assuming that the outflows had their axes along the l.o.s.~
The maximum values were derived using the maximum mass estimates 
and assuming a mean inclination angle i$\sim57.3^{\circ}$ to account for inclination effects 
(leading to a mean correction factor $\langle f(i)\rangle = 2.9$ for the momentum flux -- cf. \citealt{bontemps1996}).

We stress that the amount of outflow mass hidden at low 
velocities in the main body of the $^{12}$CO(2--1) line, M$_{\rm{low}}$, 
contributes only $\sim$~1$\%$ to 5$\%$ of the maximum momentum flux computed here, 
due to the low velocity of this material. 
Therefore, thanks to the procedure adopted here to take into account 
the different velocity regimes of the outflows, 
the relatively large uncertainty in the outflow-entrained mass at low velocities results 
only in a small uncertainty factor for the momentum flux.

The main source of uncertainty in our momentum and momentum flux estimates comes from the hypothesis made on the inclination angle with respect to the l.o.s. .
We considered that our sample of eleven outflow lobes was large enough to 
justify a statistical treatment and applied a mean inclination angle of i$\sim$57.3$^{\circ}$ 
assuming random outflow orientations.
We estimate that this hypothesis on the inclination angle leads to a factor of $\sim 2$ 
uncertainty on the individual momentum and momentum flux values.
Altogether the global uncertainty on the derived momentum fluxes is roughly a factor 
of $\sim 8$, which is consistent with the results of  \citet{cabrit1990} based 
on an analysis of synthetic outflow data.

Table~\ref{tab:summary_tab} gives the total dynamical parameters of the entire population 
of outflows observed in NGC~2264-C obtained by summing the values computed 
for the eleven outflow lobes (from Table~\ref{tab:big_tab}). 
The total kinetic energy contained in the eleven outflows was estimated from the total momentum as follows:
E$_{\rm{tot}}$= $\sum \limits_{flows}$1/2 $\times$ P$_{\rm{flow}}$ $\times$ V$_{\rm{char}} \sim$  3.4$\times$10$^{44}$ -- 1$\times$10$^{45}$ ergs.
We also estimated the rate of kinetic energy deposited by the outflows in the surrounding medium as: 
L$_{\rm{tot}}$= 1/2 $\times$ F$_{\rm{tot}}$ $\times$ $\rm{V}_{\rm{esc}}$, 
where $\rm{V}_{\rm{esc}}$ is the escape velocity of the cloud 
$\rm{V}_{\rm{esc}}$=$\sqrt{2GM_{\rm{cloud}}/R_{\rm{cloud}}}$ $\sim$ 5 km.s$^{-1}$ 
\citep{stojimirovic2006}. 
In this way, we found a total mechanical luminosity L$_{\rm{tot}}\sim$ 0.21 -- 1.2~L$\bm{_{\odot}}$.
\newline The range of values we derive for the total mass of outflow-entrained gas 
is in agreement with the mass values found
by \citet{margulis1988} for the whole NGC~2264-C region. 
The range of values we find for the overall momentum and mechanical luminosity are in agreement only with the lower values
of \citet{margulis1988}. 
This difference is mainly due to their method, which used the maximum projected flow velocity 
to compute upper limits to the momentum and mechanical luminosity. 
Furthermore, \citet{margulis1988} did not resolve the network of outflows in NGC~2264-C 
and thus used a single characteristic velocity for the eleven outflows, while our study reveals 
large velocity variations from outflow to outflow.

\begin{table*}[!ht]
\begin {center}
\centering \par \caption{Global properties of the population of eleven $^{12}$CO(2--1) outflows observed in NGC~2264-C.}
\begin{tabular}{ccccc}
\hline
\hline
 {M$_{\rm{tot}}$} & {P$_{\rm{tot}}$} &  {F$_{\rm{tot}}$} & {E$_{\rm{tot}}$} & {L$_{\rm{tot}}$} \\
 {($10^{-2}$ M$\bm{_\odot}$)} & {(M$\bm{_\odot}$.km.s$^{-1}$)} 
 & {($10^{-5}$ M$\bm{_\odot}$.km.s$^{-1}$.yr$^{-1}$)} & {(10$^{44}$ ergs)} & {(L$\bm{_{\odot}}$)} \\
 $^{(1)}$ & $^{(2)}$ & $^{(3)}$ & $^{(4)}$ & $^{(5)}$ \\
\hline
 \hfill & \hfill & \hfill & \hfill & \hfill \\
 3 -- 189 & 3.8 -- 11 & 30 -- 170 & 3.4 -- 10 & 0.2 -- 1.2  \\
  \hfill & \hfill & \hfill & \hfill & \hfill \\
       \hline
\end{tabular}
\vspace*{-0.45ex}
\begin{list}{}{}
\item[$^{(1)}$]{Minimum and maximum total mass of outflow-entrained gas, computed from the individual outflow masses given in Table~\ref{tab:big_tab}.}
\item[$^{(2)}$]{Minimum and maximum total outflow momentum, computed from the individual outflow momenta given in Table~\ref{tab:big_tab} (P$_{\rm{tot}}$ =$\sum \limits_{flows}$ P$_{\rm{flow}}$).}
\item[$^{(3)}$]{Minimum and maximum total outflow momentum flux (F$_{\rm{tot}}$ =$\sum \limits_{flows}$ F$_{\rm{flow}}$).}
 \item[$^{(4)}$]{Minimum and maximum total kinetic energy of the outflows.}
 \item[$^{(5)}$]{Minimum and maximum total mechanical luminosity of the outflow population.}
 \end{list}
\label{tab:summary_tab}
\end {center}
\end{table*}

\section{Discussion}

\subsection{Identification of driving sources}
~~~ The lobes named F1 and F2 in Fig.~\ref{fig:allmap} are the most elongated and best collimated 
outflow lobes identified in our study. 
F1 extends over 130\arcsec ~and F2 extends over 180\arcsec, corresponding to 
physical lengths of 0.49~pc and 0.71~pc, respectively, assuming the outflows are perpendicular to the line of sight (i$=90^\circ$).
\begin{figure}[h]
\centering
\includegraphics[width= 0.9\linewidth,angle=-90]{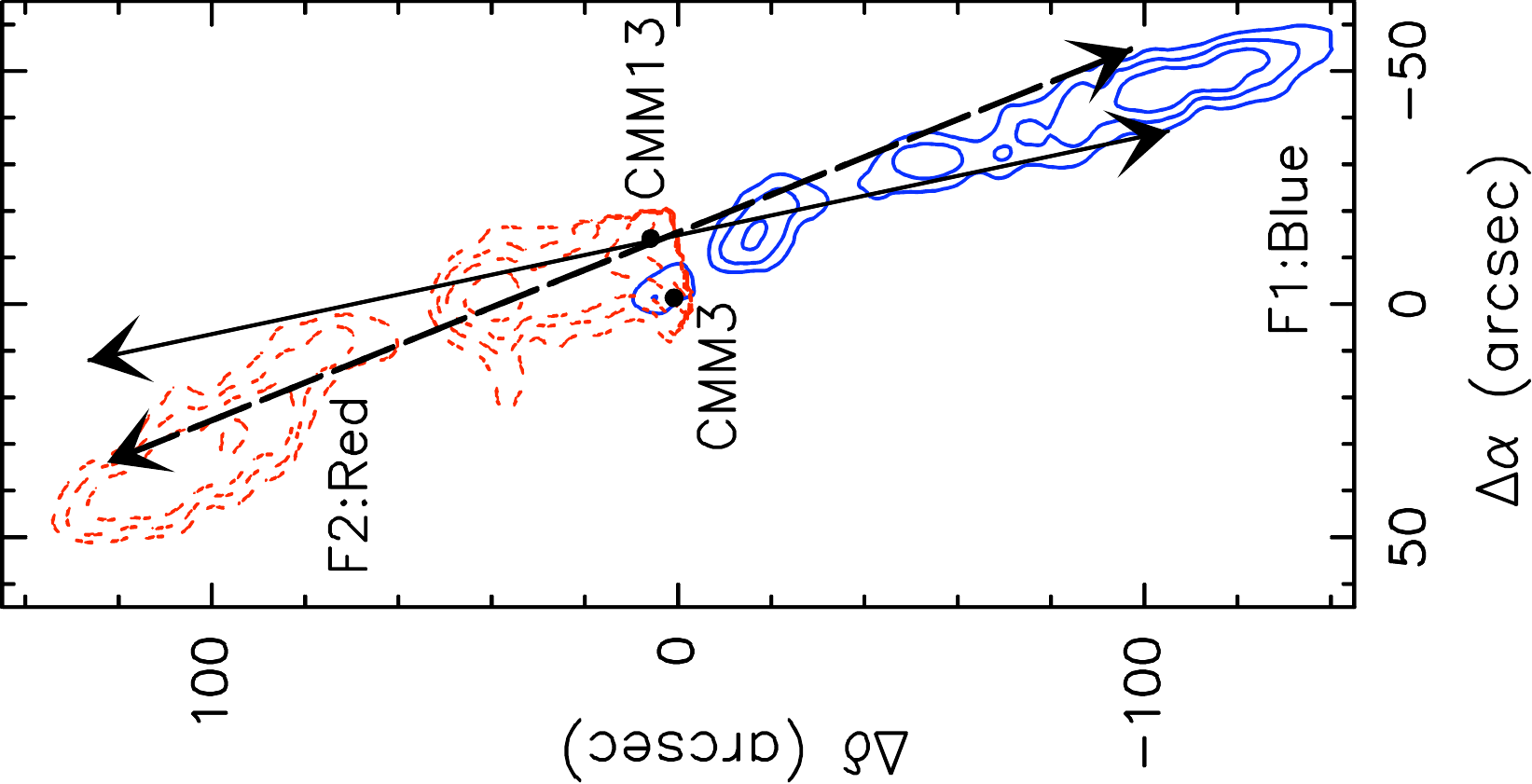}\\
\caption{Blow up of the $^{12}$CO(2--1) map of Fig.~\ref{fig:allmap} toward the (F1, F2) bipolar outflow in  
the central part of the NGC~2264-C protocluster. The full contours show the integrated intensity in the blue-shifted part of CO(2--1) line (from 1 to $-27$ km/s). Contours range from 5 to 98 K.km.s$^{-1}$. The dashed contours are levels of CO(2--1) integrated intensity in the red-shifted part of the line (from 13 to 28.5 km/s), and contours range from 5 to  110 K.km.s$^{-1}$. Markers refer to the positions of millimeter continuum peaks C-MM3 and C-MM13. 
The solid line arrow marks the direction of the bipolar outflow made up by F1 and F2, at a position angle of $12^{\circ}$ (east of north). 
The dashed line arrow marks the direction of the (F1, F2) bipolar outflow, at a position angle of $22^{\circ}$. 
This corresponds to the range of outflow axes for which the most symmetric PV diagrams are obtained (see Fig.~\ref{fig:posvelo}).
}
\label{fig:outflow_axis}
\end{figure}
%
The similar morphologies and relative positions of the two lobes in the map of Fig.~\ref{fig:allmap}
strongly suggest that F1 and F2 are physically associated and correspond to the two lobes of a single bipolar outflow. 
Based on Fig. \ref{fig:allmap} and Fig.~\ref{fig:outflow_axis}, the best candidate driving sources for this bipolar flow are C-MM3 and C-MM13. 

\begin{figure}[!h]
\centering
\includegraphics[width=0.9\columnwidth,angle=-90]{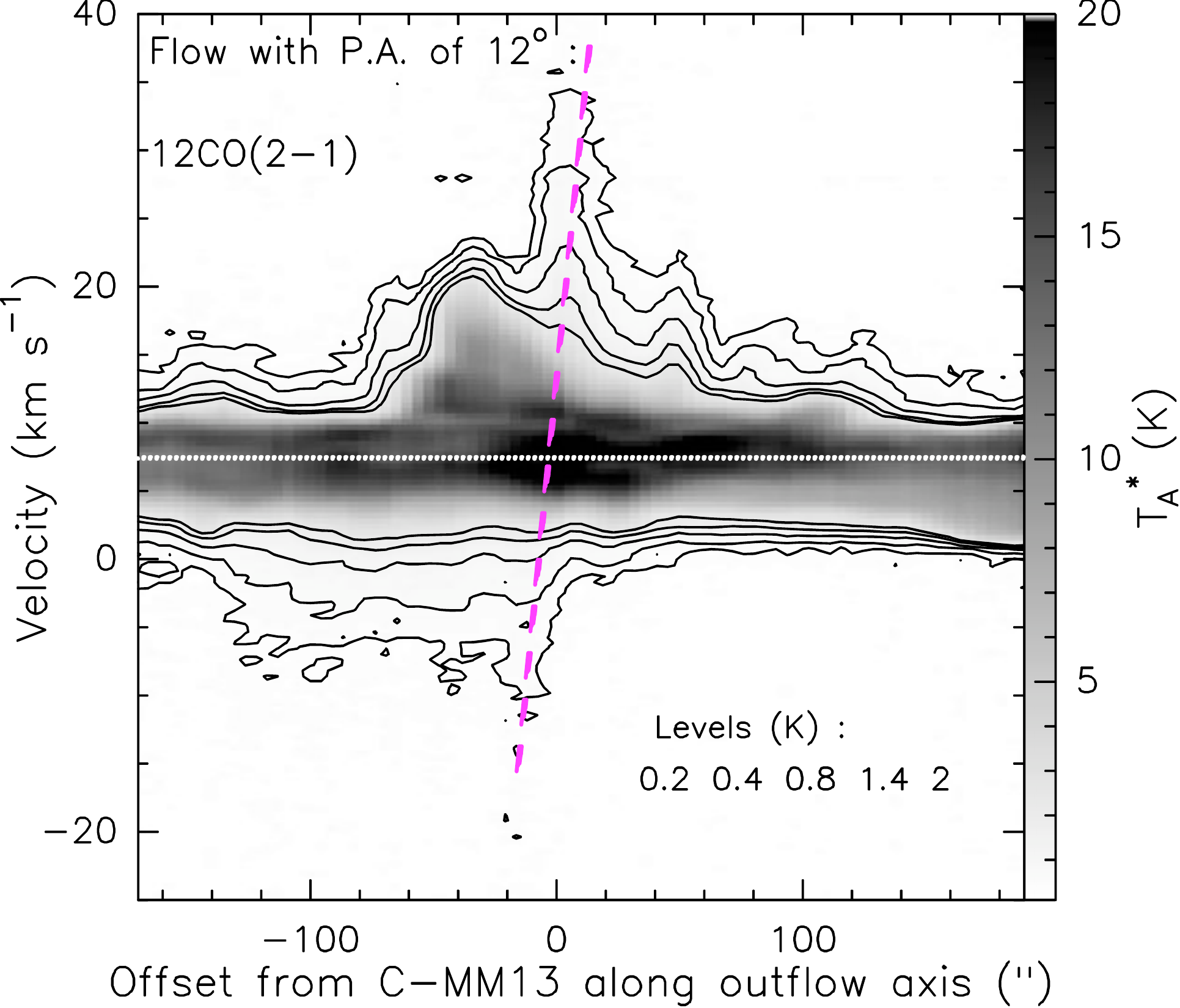}\\
\caption{Position-velocity diagram of the (F1, F2) bipolar flow, constructed along an axis passing through C-MM13 
(offset $0\arcsec$) and defined by a position angle of $12^{\circ}$ in the plane of the sky, and oriented from south to north (see Fig.~\ref{fig:outflow_axis}).
The dotted horizontal line represents the value of the systemic velocity observed at the center of the protocluster.
The dashed line in red represents the best symmetry axis for the highest velocity emission.
Note that most of the red-shifted velocity occurring at negative offsets (i.e. southern) is due to outflow F11 and is unrelated to (F1,F2). 
If we ignore this F11 emission feature, the red-shifted and blue-shifted high-velocity components seen in 
the diagram exhibit a symmetric pattern with respect to offset $0\arcsec $ 
(intersection of the dotted and dashed lines), which corresponds to the position of the protostellar core C-MM13.}
\label{fig:posvelo}
\end{figure}

To determine which source is driving the (F1, F2) bipolar outflow, we constructed several 
position-velocity (PV) diagrams along the axis of the flow, assuming various barycenters and 
orientations for the outflow. 
Figure~\ref{fig:posvelo} shows the best (most symmetric) position-velocity diagram we found. 
This PV diagram was constructed along an outflow axis passing through C-MM13 and defined 
by a position angle of $12^\circ $ (east of north) on the plane of the sky.
Note that the PV diagrams constructed for position angles between $12^\circ $ and $22^\circ $ are nearly indistinguishable.
Figure~\ref{fig:outflow_axis} indicates the range of possible outflow axes based on our PV-diagram study.
It is clear from Fig.~\ref{fig:posvelo} that the highest-velocity emission ($\pm 20$~km/s with respect 
to the systemic velocity) is detected in the immediate vicinity of the protostellar core C-MM13, 
and that this high-velocity emission presents a very high level of  symmetry  about C-MM13. 
On this basis, we conclude that C-MM13 is the most likely driving source of the bipolar outflow made up by lobes F1 and F2.

One possible explanation for the slight misalignment between the outflow features located close to C-MM13 and their more distant 
counterparts (see Fig.~\ref{fig:outflow_axis}) is that the protostellar jet driving the (F1, F2) outflow precesses with time \citep{gueth1998}. 
High-velocity features located farther away from the driving source, which presumably correspond to older ejection events,  
would then indicate a different outflow direction than that indicated by the high-velocity features located closer to C-MM13.

Based on spatial coincidence and absence of confusion (see Fig.~\ref{fig:allmap}), we also believe that the blue-shifted lobe F3 is driven by 
the millimeter source C-MM2 and that the red-shifted lobe F4 is related to C-MM1. 
The seven other lobes are either too poorly collimated (eg. red-shifted lobe F11), too far away from the protostellar clump (eg. red-shifted lobe F6), 
or too blended with other lobes to be directly associated with any particular Class~0-like object. 
\\~~

The outflow lobe F4  exhibits a compact ($\sim$ 0.03 pc) $^{12}$CO 
emission peak at v$_{LSR} \sim$ +18 km.s$^{-1}$, corresponding to 
a projected velocity offset $\sim$ +11 km.s$^{-1}$ from the systemic velocity (see Fig.~\ref{fig:bound_F4}). 
No emission is seen at v$_{LSR} =$ +18 km.s$^{-1}$ in our $^{13}$CO data, ruling out a background cloud origin for 
the  $^{12}$CO peak. Such a discrete $^{12}$CO emission feature, well defined in both space and velocity, is 
reminiscent of the high-velocity molecular  ``bullets" observed toward some young outflows driven by Class~0 objects 
such as the L1448-C bipolar flow \citep{bachiller1990}.

We also stress that there is evidence for at least two well-collimated outflows in NGC~2264-C: 
the first one is the (F1, F2) pair which  is very likely driven by  CMM-13, while the second one is the outflow lobe F10 which 
is also well collimated but of unidentified origin.  
Given that the best collimated CO outflows are generally detected toward the youngest protostellar objects (\citealt{andre1990}, \citealt{bachiller1996}), 
these highly-collimated outflows further testify to the presence of extremely young (Class~0) protostellar objects 
in the NGC~2264-C protocluster. 

\subsection{Properties of the driving sources and comparison with well-studied outflows}

We now discuss the properties of the three protostellar driving sources identified in the previous section : C-MM1, C-MM2, and C-MM13.

\subsubsection{Source properties}
We used {\it{Spitzer}}/MIPS data at 70 $\mu$m, APEX/P-ArT\'eMiS data at 450 $\mu$m (see Fig.~\ref{fig:cont}b), 
and IRAM~30m MAMBO data at 1.2~mm
\citep{peretto2006} to estimate the bolometric luminosities (L$_{\rm{bol}}$) of C-MM1, C-MM2 and C-MM13. 
We could directly measure 70 $\mu$m and 450 $\mu$m flux densities for the relatively isolated sources C-MM1 and C-MM2 and we then estimated their bolometric luminosities using the online tool devised by \citet{robitaille2007} to fit the spectral energy distributions of young stellar objects (YSOs). 
Unfortunately, due to the presence of the bright young star IRS1 (see Fig.~\ref{fig:cont}) in the field, it was impossible to derive a flux from the {\it{Spitzer}}/MIPS data for C-MM13. 
We could nevertheless obtain a MIPS 70 $\mu$m flux density and a bolometric luminosity for the neighboring  source C-MM3 ($\sim$13.5$\arcsec $ away from C-MM13).  
We then derived a luminosity estimate for C-MM13 assuming that the bolometric luminosity ratio L$^{C-MM13}_{\rm{bol}}$/L$^{C-MM3}_{\rm{bol}}$ was 
equal to the 450 $\mu$m flux density ratio F$^{C-MM13}_{450\mu m}$/F$^{C-MM3}_{450\mu m}$. 
We stress that the bolometric luminosity derived in this way for C-MM13 is more uncertain than the $L_{\rm{bol}}$ estimates obtained for C-MM1 and  C-MM2. 

\noindent Our  bolometric luminosity estimates for the three identified driving sources are listed in Table~\ref{tab:Lbol_Menv_tab}.
We also give the envelope masses (M$_{\rm{env}}$) of these protostellar objects as measured by \citet{peretto2007} with the IRAM Plateau de Bure interferometer on $\sim$3000 AU scales. 

\begin{table}[!h]
\begin {center}
\centering \par \caption{Properties of the sources : C-MM1, C-MM2, C-MM13, and C-MM3.}
\begin{tabular}{ccccc} 
\hline
\hline
{Core} & {Outflow} & {M$_{\rm{env}}$} & {L$_{\rm{bol}}$} & {F$_{\rm{flow}}$} \\ 
{} & {lobe} & {(M$_{\odot}$)} & {(L$_{\odot}$)} & {(10$^{-5}$M$_{\odot}$.km.s$^{-1}$.yr$^{-1}$)}\\ 
\hline
{C-MM1} & F4 &{6.2 $^{+6.2}_{-3.1}$} & {7 $\pm$ 3} & {0.5 -- 3.2}\\
{C-MM2} & F3 &{5.5 $^{+5.5}_{-2.7}$} & {11 $\pm$ 3} & {6.3 -- 28.3}\\ 
{C-MM13}& (F1,F2) &{7.6 $^{+7.6}_{-3.8}$} & {8.6 $\pm$ 5} & {4.2 -- 16.6}\\ 
{C-MM3} & - &{15.2 $^{+15.2}_{-7.6}$} & {50 $\pm$ 10} & {-}\\
 \hline 
\end{tabular}
 \label{tab:Lbol_Menv_tab} 
 \end {center}
\end{table} 

\begin{figure}[!h]
   \begin{center}
\includegraphics[width= 0.8\columnwidth,angle=-90,trim=0cm 0.5cm 0cm 0cm, clip=true]{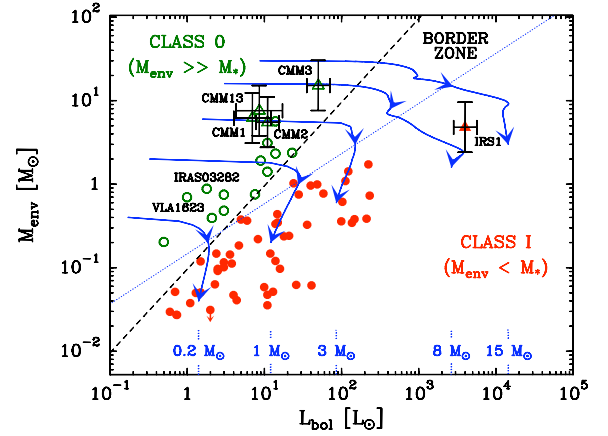}
\caption{Envelope mass versus bolometric luminosity diagram comparing the locations of the protostellar sources of NGC~2264-C (open triangles with error bars) with the positions of low-mass Class~I (filled circles) and Class~0 objects (open circles) (from \citealt{andre2000}). Model evolutionary tracks, computed for various final stellar masses M$_{\star}^{final}$, are shown.
The dashed and dotted lines are two M$_{env}$ -- L$_{\rm{bol}}$ relations marking the conceptual border zone 
between the Class 0 (M$_{\rm{env}}$ $>$ M$_{\star} / \epsilon$) and the Class I 
(M$_{\rm{env}}$ $<$ M$_{\star} / \epsilon$) stages, where $\epsilon \approx 50\%$ is 
the local star formation efficiency (see \citealt{andre2008} for details).}
\label{fig:Menv_Lbol}
\end{center}
\end{figure}

The protostellar sources C-MM1, C-MM2, C-MM3 and C-MM13 were then placed in an $M_{\rm{env}}$--$L_{\rm{bol}}$ diagram 
(see Fig.~\ref{fig:Menv_Lbol}) similar to the ones presented by \citet{andre2000,andre2008}. 
Model evolutionary tracks are shown, computed by assuming each protostar 
forms from a bounded condensation of finite initial mass $M_{\rm{env}}$(0) and has 
$L_{\rm{bol}}=GM_{\star}(t)\dot{M}_{\rm{acc}}(t) / R_{\star}(t) + L_{\star}(t)$, where $R_{\star}$ is the 
protostellar radius and $L_{\star}$ the interior stellar luminosity \citep{stahler1988,hosowaka2008}. 
The mass accretion rate and the envelope mass were assumed to be related by 
$\dot{M}_{\rm{acc}}(t) = \epsilon M_{\rm{env}}(t) / \tau$, where $\epsilon=50\%$ is the typical 
star formation efficiency for individual cores \citep{matzner2000}, and $\tau = 10^{5}$ yr 
is the characteristic timescale of protostellar evolution, leading to 
$\dot{M}_{\rm{acc}}(t)$ and $M_{\rm{env}}(t)$ functions declining exponentially with time 
(see \citealt{bontemps1996}). The positions of C-MM1, C-MM2, C-MM3 and C-MM13 
in this diagram demonstrate that they are bona fide Class~0 objects.
Moreover, comparison with the model evolutionary tracks suggests that these objects
will form stars with main-sequence masses M$_{\star}^{final} \sim$ 2 - 8~ M$_{\odot}$.

\subsubsection{Comparison with other outflows from the literature}

Here, we compare the properties of the three outflows (F1, F2), F3, and F4 with those of several protostellar outflows from the literature. 

\begin{figure*}[!ht]
\begin{center}
  \subfigure{\includegraphics[width=0.8\columnwidth,angle=-90,trim=0cm 0cm 0cm 29cm, clip=true]{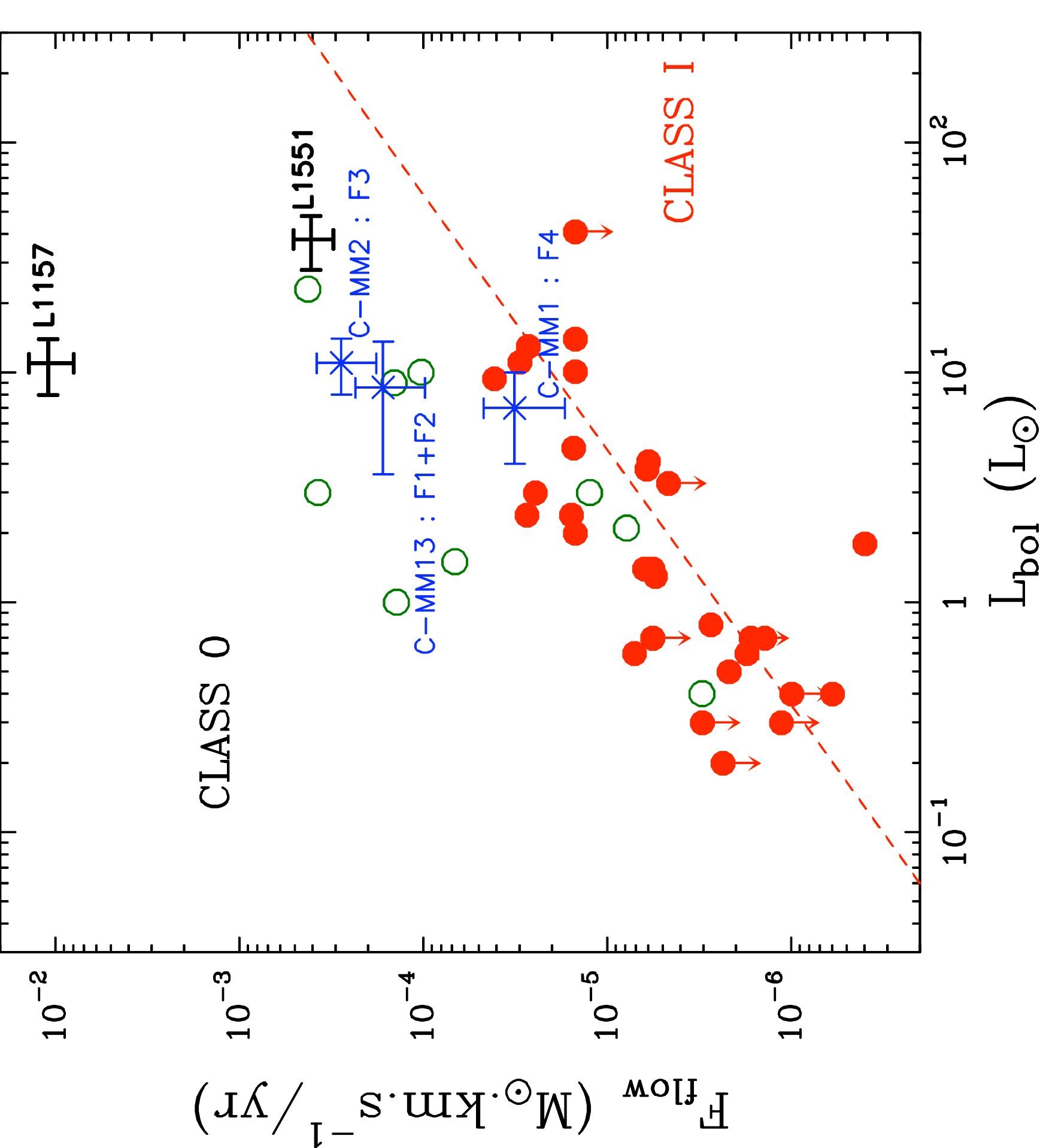}}
  \hfill
  \subfigure{\includegraphics[width=0.84\columnwidth,angle=-90,trim=0cm 0cm 0cm 33cm, clip=true]{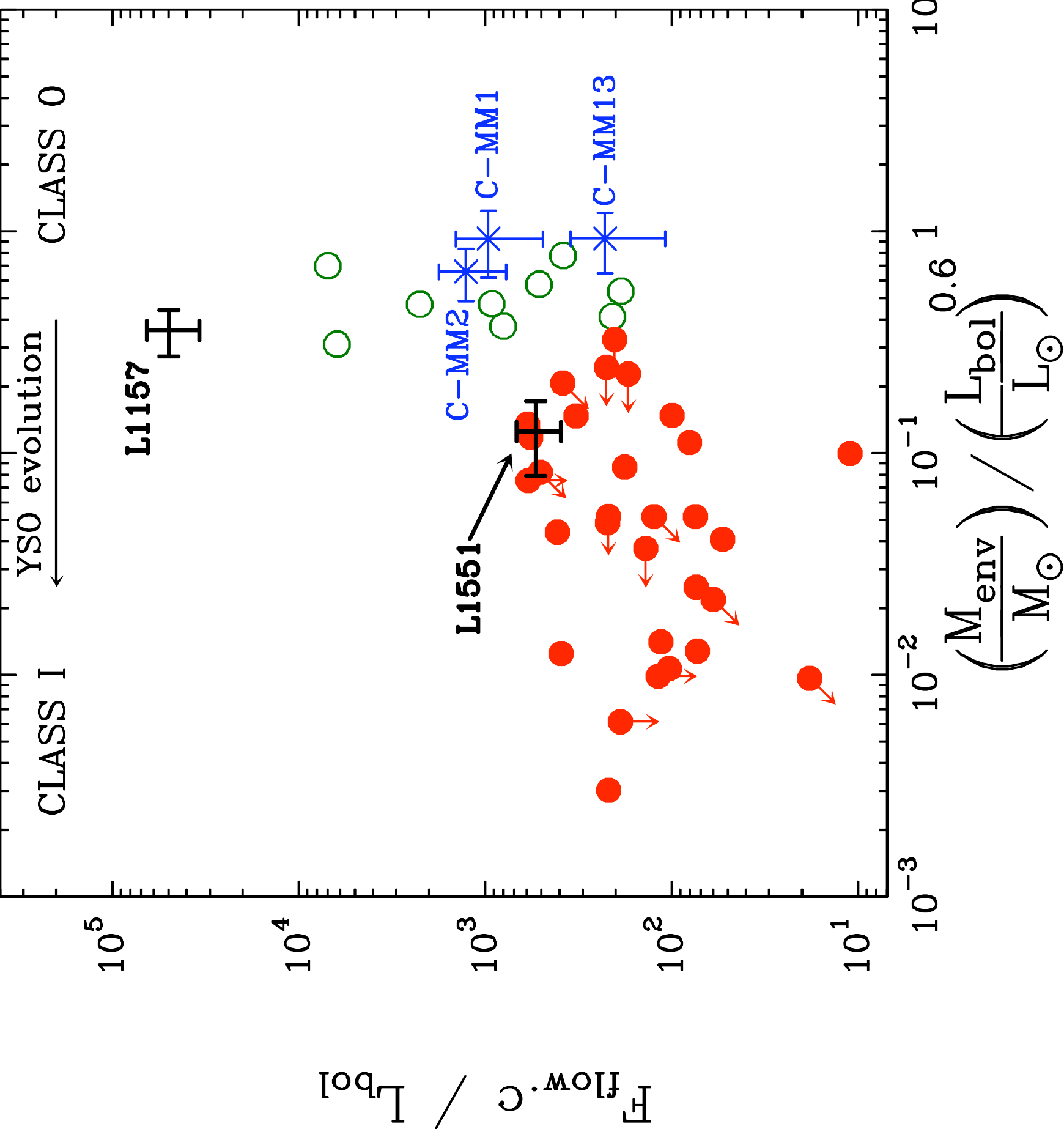}}
  \caption{CO momentum flux F$_{\rm{flow}}$ versus bolometric luminosity (left), and F$_{\rm{flow}}$c/L versus M$_{\rm{env}}$/L$_{\rm{bol}}$$^{0.6}$ (right).
The three candidate Class~0 objects identified as the driving sources of the outflows (F1, F2), F3 and F4 in NGC~2264-C are shown as blue stars. For comparison, the Class 0 and Class I objects studied by \citet{bontemps1996} are shown as open and filled circles, respectively. 
The driving sources of the two molecular outflows taken as references by \citet{nakamura2007} are shown by black error bars.}
\label{fig:Fco_vs_Lbol}
\end{center}
\end{figure*}

\citet{bontemps1996} carried out a study of outflow energetics for a sample of 45 low-luminosity YSOs including both Class~I and Class~0 objects. 
They found that outflows from younger sources were more powerful.
The outflow momentum flux values we find here, namely a few 10$^{-5}$ M$_\odot$.km.s$^{-1}$.yr$^{-1}$, are very similar 
to the momentum fluxes of the Class~0 outflows in the Bontemps et al. sample (see Fig.~\ref{fig:Fco_vs_Lbol}).
The left panel of Fig.~\ref{fig:Fco_vs_Lbol} plots outflow momentum flux versus driving source luminosity, while 
the right panel shows F$_{\rm{flow}}$c/L versus M$_{\rm{env}}$/L$_{\rm{bol}}$$^{0.6}$ (see \citet{bontemps1996} for details).
Figure \ref{fig:Fco_vs_Lbol} 
shows that  C-MM1, C-MM2, and C-MM13 drive outflows which are typical of low- to intermediate-mass Class~0 objects 
and closely follow the trend established by \citet{bontemps1996}.

We can also compare the properties of the outflows mapped in NGC~2264-C with the 
two molecular outflows used as references by \citet{nakamura2007} to estimate the 
typical outflow momentum to be injected in their numerical simulations.
\newline The first one is the well-documented outflow driven by the border-line Class~I/Class~0 object L1551 IRS~5. 
\citet{stojimirovic2006} showed that the L1551 IRS~5 bipolar outflow extends over 1.3~pc and drives a total 
momentum of $\sim$~17.8 -- 23 M$_{\odot}$.km.s$^{-1}$.
The dynamical parameters of the L1551 outflow, as re-estimated by us from the results of 
\citet{stojimirovic2006} for consistency with the present study, 
are reported in Table~\ref{tab:big_tab} (see also Fig.~\ref{fig:Fco_vs_Lbol}).
We note that the momentum flux injected by the L1551 outflow in the surrounding medium is of the same order as 
the momentum flux of the strongest outflow observed in the NGC~2264-C protocluster (F11).
\newline The second reference outflow used by \citet{nakamura2007} is the one driven by the 
low-luminosity Class~0 protostar L1157-MM \citep[e.g.][]{umemoto1992, bachiller2001}.
The dynamical parameters of the L1157 outflow are also reported in Table~\ref{tab:big_tab} and used in Fig.~\ref{fig:Fco_vs_Lbol}.
It can be seen in Fig.~\ref{fig:Fco_vs_Lbol} that the L1157 outflow is characterized by a very high momentum flux. 
None of the outflows mapped in the NGC~2264-C protocluster drives that much momentum flux in 
the surrounding medium, and even the combined effects of the eleven outflows in NGC~2264-C inject at most 
F$_{\rm{tot}} \sim$~1.7$\times$10$^{-3}$~M$_{\odot}$.km.s$^{-1}$.yr$^{-1}$ (see Table~\ref{tab:summary_tab}), i.e one tenth of what is found for the L1157 outflow alone. 
Therefore, we conclude that L1157-MM  represents an extreme case of a Class~0 outflow.

We also note that the total momentum flux derived here (Table~\ref{tab:summary_tab}) for the network of 
eleven outflows in NGC~2264-C is similar to the total outflow momentum flux found by \citet{knee2000} in the NGC~1333 protocluster.

\subsection{Outflows as a source of turbulence and support}

\subsubsection{Observed turbulence vs outflow contribution}
We now discuss whether or not the eleven outflows mapped in NGC~2264-C are likely to contribute a significant fraction 
of the turbulence observed in the protocluster. 

From our $^{13}$CO mapping, we determined a one-dimensional velocity dispersion 
$\sigma_{\rm{v}}$ 
$\sim$ 1.7 km.s$^{-1}$, measured on the mean $^{13}$CO(2--1) spectrum over the NGC~2264-C clump (radius $R \sim 0.7$~pc). 
If we place NGC~2264-C in the linewidth-size diagram representing Larson's first scaling law  \citep{larson1981}, we find that 
NGC~2264-C exhibits a velocity dispersion about twice as large as what is expected from the classical linewidth-size relationship 
observed in giant molecular clouds (GMCs) \citep{solomon1987}. This suggests that NGC~2264-C is more turbulent 
than the CO clouds considered by \citet{solomon1987}.
However, \citet{heyer2008} recently re-examined Solomon's results with new $^{13}$CO data and 
found the relationship $\sigma_{\rm{v}}$ = $\sqrt{\pi G \Sigma / 5} \times R^{1/2}$, which shows that 
the normalization of Larson's first scaling law depends on the mean surface density $\Sigma$ of the clouds. 
From the 1.2 mm dust continuum observations of \citet{peretto2006} we estimate a mean surface density of $\sim 3000$~M$_{\odot}$.pc$^{-2}$ 
for the NGC~2264-C cluster-forming clump, which is much higher than the mean surface density $\sim 170$~M$_{\odot}$ pc$^{-2}$ 
of the CO clouds following the classical linewidth-size relation.
The location of NGC~2264-C in the $\sigma_{\rm{v}}$ / R$^{0.5}$ = f($\Sigma$)  diagram  of \citet{heyer2008} 
is nevertheless similar to that of the densest clouds discussed by these authors.  
We also note that  both the N$_{2}$H$^{+}$ \citep{peretto2006} and $^{13}$CO (this work)
linewidths of NGC~2264-C are consistent with the log($\Delta$v) = 0.23 + 0.21$\times$log(R) correlation 
found by \citet{caselli1995} for massive cloud cores in Orion.
Therefore, although NGC~2264-C exhibits broader linewidths than those expected 
from Larson's classical linewidth-size relation, it exhibits the same level of turbulence as that generally 
found in massive clouds and cores. 

Next, we can estimate the rate of turbulent energy dissipation in the NGC~2264-C clump 
using equation~(7) of \citet{maclow1999}. 
From the 1.2~mm dust continuum map obtained by \citet{peretto2006}, a total gas mass of 2300~M$_{\odot}$ 
can be derived within a radius of 0.7~pc.
Using the velocity dispersion derived from our $^{13}$CO observations,  we find : 
\begin{equation}
\rm{L}_{\rm{turb}} \sim {\displaystyle{ \frac{1/2 \times \rm{M} \sigma_{\rm{v}}^{2}}{\rm{R} / \sigma_{\rm{v}}}}} \sim 1.2~L_{\odot}. 
\end{equation}
For comparison, the total mechanical power due to the eleven outflows in NGC~2264-C 
is {\large{$\rm{L}_{\rm{tot}}$ = 0.7 $\pm$ 0.5~L$_{\odot}$}} (see \S~4.2). 
We conclude that  the network of protostellar outflows observed 
in NGC~2264-C can contribute a very significant, if not dominant, 
fraction of the turbulence observed in the protocluster.

\subsubsection{Gravitational collapse vs outflow support}
In order to determine whether the force exerted by the outflows on NGC~2264-C can slow down or even stop the global collapse of the protocluster, 
we first estimate the force needed to keep the whole protocluster in hydrostatic equilibrium. 
\newline\noindent The pressure gradient needed to balance the gravitational acceleration  
$a_{\rm{grav}}(r) = G M(r) / r^{2}$ at radius $r$ 
and keep a spherical clump with a mass distribution $M(r)$  in hydrostatic equilibrium is:
\begin{equation}
\frac{dP_{\rm{grav}}}{dr} =  - \rm{G}  \frac{\rho(r).M(r)}{r^{2}}
\end{equation}
Assuming a density distribution of the form $\rho$(r)~=~{\large{$\frac{\sigma^{2}}{2\pi Gr^{2}}$}}, 
this corresponds to a supporting pressure 
$P_{\rm{grav}}(R) = G \times M(R)^{2} / 8\pi R^{4}$ for the spherical shell of radius $R$ 
enclosing a mass \\$M(R)=2\sigma^{2} R / \rm{G}$.
\newline Therefore, the total force needed to balance gravity at radius $R$ in the NGC~2264-C protocluster is:
\begin{equation}
F_{\rm{grav}}(R) = P_{\rm{grav}}(R) \times 4 \pi R^{2} = G \times M(R)^{2} / 2R^{2} \\
\end{equation}
The corresponding gravitational binding energy is 
\begin{equation} 
E_{\rm{grav}}(R) = - G \times M(R)^{2} / R
\end{equation}
If we adopt $R_{1}$= 0.4~pc and $M(R_{1})$ = 1600 M$_{\odot}$, we find : \\
$\rm{F}_{\rm{grav}}$ = 40$\times$10$^{-3}$~M$_{\odot}$.km.s$^{-1}$.yr$^{-1}$, \\and 
a$_{\rm{grav}}$ = 50~km.s$^{-1}$.Myr$^{-1} \sim$ 50~pc.Myr$^{-2}$.
\\ If we adopt $R_{2}$= 0.7~pc and $M(R_{2})$ = 2300 M$_{\odot}$, we find :  \\
$\rm{F}_{\rm{grav}}$ = 23$\times$10$^{-3}$~M$_{\odot}$.km.s$^{-1}$.yr$^{-1}$, \\and 
$a_{\rm{grav}}$ = 20~km.s$^{-1}$.Myr$^{-1} \sim$ 20~pc.Myr$^{-2}$.
\\ ~~

Comparing {\large{F$_{\rm{grav}}$}} to our upper estimate of the force exerted by outflows on the NGC~2264-C protocluster, 
{\large{F$_{\rm{tot}}$}} = 1.7$\times$10$^{-3}$M$_{\odot}$.km.s$^{-1}$.yr$^{-1}$ (see \S4.2), 
we find that the eleven observed outflows fall short by a factor $\sim 25$ to 
provide significant support against collapse in NGC~2264-C. 

We also note that the turbulent acceleration \citep[cf.][]{matzner2007} : 
\begin{equation}
a_{\rm{turb}} = \vert \frac
{\vec{\nabla (\rho \sigma_{\rm{v}}^2)}}{\rho} \vert \sim \sigma_{\rm{v}}^{2} / R
\end{equation} 
is only $a_{\rm{turb}} \sim$ 4.7~km.s$^{-1}$.Myr$^{-1}$ at radius R~$= 0.7$~pc, 
which is $\sim$ 4-5 times lower than the gravitational acceleration derived at the same radius. 
Therefore, the enhanced level of turbulence observed in NGC~2264-C (cf. \S~5.3.1)
is also insufficient to support the clump against global collapse.

From an energetic point of view, the total kinetic energy (see E$_{\rm{tot}}$ in Table~\ref{tab:summary_tab}) derived in \S4.2 for the eleven outflows 
is nearly three orders of magnitude lower than the gravitational binding energy of the NGC~2264-C cluster-forming clump (E$_{\rm{grav}} \sim$ 5.5$\times$10$^{47}$ ergs).
\citet{stojimirovic2006} showed that the kinetic energy injected by the L1551-IRS5 outflow in the surrounding medium was 
comparable to the binding energy of the L1551 cloud itself. 
The kinetic energy they derived from the L1551 outflow (1.5$\times$10$^{45}$ ergs) is comparable to our upper estimate 
of the total kinetic energy of the outflow population in NGC~2264-C (see Table~\ref{tab:summary_tab}), but the L1551 cloud is much less massive (110~M$_{\odot}$) 
than NGC~2264-C for an equivalent volume (radius of 0.5~pc). 
Likewise, we find that the total mass of outflow-entrained material represents only 0.25$\%$ of the mass 
of the NGC~2264-C protocluster, whereas the IRS5 outflow represents $\sim$5$\%$ of the total mass of the L1551 cloud.

For comparison, the rate of gravitational energy release due to global collapse is :
\begin{equation}
L_{\rm{grav}} = -\frac{dE_{grav}}{dt} = -\left( \rm{G} M(R)^{2} / R^2 \right) \dot{r}  = \vert \rm{E}_{grav}  \vert \, 
 \frac{\rm{V}_{\rm{infall}}}{R} \sim 15\, \rm{L}_{\odot},
\end{equation}
 in the NGC~2264-C, where we have used the large-scale infall velocity $\rm{V}_{\rm{infall}} = 1.3$~km/s  measured by \citet{peretto2006}.
This is a factor of $\sim $~15--20 larger than both the total mechanical power of the observed outflows (cf. \S ~4.2)  and 
the rate of turbulent energy dissipation in the clump (cf. \S ~5.3.1). 
\\~~

We conclude that neither the direct force exerted by the network of observed  protostellar outflows, nor 
the turbulence they produce, can support the cluster-forming clump against global collapse at the present time.
Therefore, the additional source of support against gravity, not  
included in the purely hydrodynamic simulations of \citet{peretto2007} but needed to 
account for the observed characteristics of NGC~2264-C (see end of  \S ~1.2), cannot find its  
origin in protostellar feedback. Although little is known about magnetic fields in the region, the most 
plausible remaining hypothesis is that the missing support is magnetic in character.
In any case, gravitational contraction seems to largely dominate the physics of the NGC~2264-C clump and may even be the main source 
of the high level of turbulence observed in the protocluster \citep[see also][]{peretto2006}.

\subsection{Caveats}

We recall that there is a factor of $\sim 8$ uncertainty on the outflow momentum flux values given in Table~\ref{tab:big_tab} 
(see \S ~4.2). This uncertainty is mainly due to the uncertain excitation temperature of the $^{12}$CO(2--1) transition 
and the unknown inclination angles of the outflows.
Here, we explore other effects which may have led us to underestimate the energy injected by outflows in the NGC~2264-C protocluster.

First, our $^{12}$CO(2--1) mapping survey may have missed some outflows, due to its limited sensitivity. 
We had a sensitivity limit of $\sim$~0.1~K in our $^{12}$CO(2--1) observations of NGC~2264-C. 
For a typical outflow with a characteristic velocity of $\pm$10~km.s$^{-1}$ (with respect to the systemic velocity) and 
a dynamical timescale of 10$^{4}$~years, this corresponds to a momentum flux sensitivity~of~$\sim 0.2 \times10^{-5}$~M$_{\odot}$.km.s$^{-1}$.yr$^{-1}$.
\newline A rough estimate indicates that 225 additional outflows as powerful as the (F1,F2) bipolar flow, 
or 4 outflows as energetic as the L1551 outflow, would be needed in order to balance gravity in NGC~2264-C (see \S5.2.2). 
The sensitivity of our $^{12}$CO(2--1) survey is such that we could not have missed such strong outflows.
Nevertheless, a large number of fainter, undetected outflows driven by low-mass objects may be present in the protocluster. 
\newline According to \citet{peretto2007}, the NGC~2264-C protocluster contains seven Class~0-like 
objects with envelope masses ranging from 4~M$_{\odot}$ to 16~M$_{\odot}$. 
Based on the M$_{\rm{env}}$--L$_{\rm{bol}}$ diagram shown in Fig.~\ref{fig:Menv_Lbol}, we estimate that 
these seven protostellar objects are the progenitors of stars with main-sequence masses in 
the range 2~M$_{\odot} \leq$~M$_{\star}^{final}$~$\leq$ 8~M$_{\odot}$ (see \S~5.2.1).  
If the population of protostellar objects in the NGC~2264-C protocluster follows the  \citet{kroupa2001} IMF, 
then we expect $\sim$ 250 protostellar objects with final stellar masses M$_{\star}^{final}$ in the range 
0.01~M$_{\odot} $\simlt~M$_{\star}^{final} \simlt$ 2~M$_{\odot}$.
Assuming these 250 objects lie in the border zone between the conceptual Class~0 and Class~I stages 
and have M$_{\rm{env}} \sim$ M$_{\star}^{final}$, we infer that $\sim$10 of them have 
envelope masses $M_{\rm{env}}$ in the range 1~M$_{\odot} \simlt$ M$_{\rm{env}}\simlt$ 2~M$_{\odot}$, 
$\sim$110 of them have 0.1~M$_{\odot} \simlt$ M$_{\rm{env}}\simlt$ 1~M$_{\odot}$,  
and $\sim$130 of them have 0.01~M$_{\odot} \simlt$ M$_{\rm{env}} \simlt$ 0.1~M$_{\odot}$ in NGC~2264-C.
\newline Following \citet{bontemps1996}, we adopt the following relation between outflow momentum flux and protostellar enveloppe mass:  
 log(F$_{\rm{flow}}$) $= -4.15 + 1.1 \times$~log(M$_{\rm{env}}$).
From this relation, we estimate that the $\sim 250$ low-mass protostellar objects mentioned earlier can generate
a total momentum flux of $\sim$3.3$\times$10$^{-3}$~M$_{\odot}$.km.s$^{-1}$.yr$^{-1}$.
Therefore, even if we account for the effect of weak, undetected outflows driven by low-mass protostellar objects, 
the total outflow momentum flux injected in NGC~2264-C is only  
F$_{\rm{max}} \sim$ 5$\times$10$^{-3}$~M$_{\odot}$.km.s$^{-1}$.yr$^{-1}$, 
which is still a factor of $\sim 8$ lower than the force needed to support the clump against global collapse.

Second, our $^{12}$CO(2--1) map of NGC~2264-C provides only a snapshot of outflow activity in the protocluster. Given that the characteristic timescale of protostellar evolution is $\tau \sim$ 10$^{5}$ years, 
we can estimate the total momentum P$_{\rm{max}}$ injected by protostellar outflows over their entire lifetime.
Bontemps et al. (1996) showed that outflow momentum flux is roughly proportional to protostellar envelope mass, i.e., F$_{\rm{flow}}$ decreases as 
M$_{\rm{env}}$ decreases and protostars evolve from Class~0 to Class~I  objects. 
Furthermore, the linear correlation F$_{\rm{flow}} \propto $~M$_{\rm{env}}$ is suggestive of a simple  
overall accretion/ejection history according to which M$_{\rm{env}}$, F$_{\rm{flow}}$, and the mass accretion rate, $\dot{\rm{M}}_{\rm{acc}}$,  
all decline exponentially with time (see Bontemps et al. 1996).
We thus computed the total momentum as: 
\begin{equation}
P_{\rm{max}} = \int_{0}^{+\infty}~\rm{F}_{max} \times e^{-t/\tau}~dt = \rm{F}_{max} \times~\tau.
\end{equation}
Using the value of the total momentum flux which includes the contribution from 
the low-mass protostellar population, we find P$_{\rm{max}}$ = 500~M$\bm{_{\odot}}$.km.s$\bm{^{-1}}$. 
If we consider a total number N$_{\rm{max}}$ = 256 of protostellar objects in NGC~2264-C (including the low-mass protostellar population) 
and adopt a mean stellar mass $<\rm{m}_\star>$ = 0.2~M$_{\odot}$ for the \citet{kroupa2001} IMF, then we find that the 
mean momentum per unit stellar mass is: P$_{\star}$ = P$_{\rm{max}}$ / (N$_{\rm{max}} \times <\rm{m}_\star >) \sim$ 10 km.s$^{-1}$. 
We conclude that, on average, the  NGC~2264-C protocluster exhibits a mean outflow momentum per stellar mass which is only 
$\sim 1/5$ of the value \citet{nakamura2007} used in their numerical simulations.

It should be also pointed out that the star formation rate within the NGC~2264-C 
clump may increase in the near future. 
Using a global mass inflow rate $\dot{\rm{M}}_{\rm{inf}} \sim$ 3$\times$10$^{-3}$~M$_{\odot}$.yr$^{-1}$ 
for the collapsing clump \citep{peretto2006}, we predict a star formation rate 
$\rm{SFR} = \epsilon \times \dot{\rm{M}}_{\rm{inf}} \sim$ 1.5$\times$10$^{-3}$~M$_{\odot}$.yr$^{-1}$,  
where $\epsilon = 50\%$ is the typical star formation efficiency for individual cores \citep{matzner2000}.
Such a star formation rate would lead to a momentum injection rate of the corresponding outflows 
$\rm{SFR} \times \rm{P}_{\star} \sim$ 15$\times$10$^{-3}$~M$\bm{_{\odot}}$.km.s$\bm{^{-1}}$.yr$\bm{^{-1}}$. 
This value remains lower than, but of the same order as, the force needed to support the clump against global collapse. 
This suggests that if NGC~2264-C undergoes a burst of star formation in the near future as 
a result of its global collapse, then feedback due to protostellar outflows 
may eventually become sufficient to slow down and halt the collapse of the clump.
 ~

\section{Summary and conclusions}

We carried out a high-resolution $^{12}$CO(2--1) survey of NGC~2264-C 
with the IRAM 30~m telescope in an effort to take a quantitative census of 
protostellar outflows in this collapsing cluster-forming clump. 
Additional mapping observations 
were also taken in $^{13}$CO(2--1) and C$^{18}$O(2--1). 
Our main results and conclusions can be summarized as follows:

\begin{enumerate}

\item We detected a network of eleven CO(2--1) outflow lobes with characteristic 
projected velocities between $\sim 10$~km/s and $\sim 30$~km/s, 
lengths between $\sim 0.2$~pc and $\sim 0.8$~pc, momenta between 
$\sim 0.1$~M$_{\odot}$.km.s$^{-1}$ and 
$\sim 3$~M$_{\odot}$.km.s$^{-1}$, and momentum fluxes 
between $\sim 0.5 \times 10^{-5}$~M$_{\odot}$.km.s$^{-1}$.yr$^{-1}$ 
$\sim 50 \times 10^{-5}$~M$_{\odot}$.km.s$^{-1}$.yr$^{-1}$. 

\item One pair of high-velocity CO(2--1) lobes corresponds to a highly-collimated bipolar 
flow, whose most likely driving source is the millimeter continuum object 
C-MM13 located very close to the center of mass of the NGC~2264-C protocluster. 
Two other CO(2--1) outflow lobes appear to be driven by the millimeter continuum objects C-MM1 and 
C-MM2. The driving sources of the seven other outflow lobes have not been identified.

\item The three outflows with identified driving sources have properties that are typical 
of the flows observed toward low- to intermediate-mass Class~0 protostars. 
Based on their positions in the M$_{\rm{env}}$ --  L$_{\rm{bol}}$ evolutionary diagram 
for protostars, we confirm that the sources C-MM1, C-MM2, C-MM3 and CMM13 are 
bona-fide Class~0 objects, which will form intermediate-mass stars with 
M$_{\star}^{final} \sim$ 2 -- 8~ M$_{\odot}$.

\item The total mechanical power associated with the network of eleven ouflows, 
$\rm{L}_{\rm{tot}}$ = 0.7 $\pm$ 0.5~L$_{\odot}$, is large enough to contribute 
a significant, if not dominant, fraction of the turbulence observed in the NGC~2264-C 
cluster-forming clump.

\item However, even if we account for weak, undetected outflows driven by low-mass protostellar objects 
and below the sensitivity limit of our CO(2--1) survey, the total momentum flux generated by the entire 
outflow population in the protocluster,  
F$_{\rm{max}} \sim$ 5$\times$10$^{-3}$~M$_{\odot}$.km.s$^{-1}$.yr$^{-1}$, 
falls short by nearly an order of magnitude to provide sufficient support against global collapse of the clump at the current epoch.
Therefore, the additional support against gravity missing in the hydrodynamic simulations of NGC~2264-C by 
\citet{peretto2007}, 
cannot arise from protostellar feedback, and most likely originates in magnetic fields.

\item We conclude that gravitational contraction largely dominates the physics of the NGC~2264-C clump at the present time 
and may even be the main source of turbulence in the protocluster. 
Given the early stage of evolution of the protocluster and its global collapse, 
it is nevertheless plausible that the star formation rate will increase in the future, 
up to a point where the feedback due to protostellar outflows becomes sufficient to halt the 
large-scale, global contraction of the protocluster.

\end{enumerate}
~
\newline {\it{Acknowledgements}}~ We thank the referee, Rafael Bachiller, for useful comments which helped us improve the clarity of the paper.   
We are grateful to the ArT\'eMiS team whose dedication made the P-ArT\'eMiS mapping of NGC~2264-C (Fig.~1b) possible as part of 
ESO program 080.C-0722 on APEX. 
The work presented in this paper was stimulated by discussions held during the Star Formation Workshop organized 
by Hsien Shang and Nagayoshi Ohashi at ASIAA-TIARA, Taiwan, in December 2005. 

\bibliography{ngc2264c_flows_rev_v3}

\end{document}